\documentclass[
    reprint,
    superscriptaddress,
    amsmath,amssymb,
    aps,
    prb,
    floatfix,
]{revtex4-2}
\usepackage{graphicx}
\usepackage{bm}
\usepackage{hyperref}
\usepackage{booktabs}
\usepackage{siunitx}
\DeclareSIUnit\torr{Torr}
\DeclareSIUnit\angstrom{\text {Å}}

\begin{document}

\title{Rhenium as a material platform for long-lived transmon qubits}

\author{Yanhao Wang}
    \email{yanhao.wang@yale.edu}
    \affiliation{Departments of Applied Physics and Physics, Yale University, New Haven, Connecticut, USA}
    \affiliation{Yale Quantum Institute, Yale University, New Haven, Connecticut, USA}
    
\author{Suhas Ganjam}
    \affiliation{Departments of Applied Physics and Physics, Yale University, New Haven, Connecticut, USA}
    \affiliation{Yale Quantum Institute, Yale University, New Haven, Connecticut, USA}
    
\author{Ishan Narra}
    \affiliation{Departments of Applied Physics and Physics, Yale University, New Haven, Connecticut, USA}
    \affiliation{Yale Quantum Institute, Yale University, New Haven, Connecticut, USA}
    
\author{Luigi Frunzio}
    \affiliation{Departments of Applied Physics and Physics, Yale University, New Haven, Connecticut, USA}
    \affiliation{Yale Quantum Institute, Yale University, New Haven, Connecticut, USA}
    
\author{Robert J. Schoelkopf}
    \email{robert.schoelkopf@yale.edu}
    \affiliation{Departments of Applied Physics and Physics, Yale University, New Haven, Connecticut, USA}
    \affiliation{Yale Quantum Institute, Yale University, New Haven, Connecticut, USA}

\date{\today}

\begin{abstract}
Dielectric loss at the interfaces of superconducting films has long been recognized as limiting the performance of state-of-the-art superconducting circuits. Notably, the presence of a native oxide layer on the film is hypothesized to contribute to dielectric loss at the metal-air interface. Here, we explore rhenium as a candidate for the film, motivated by its remarkable property to suppress native oxide formation. We demonstrate rhenium on sapphire as a promising material platform for superconducting circuits through the realization of transmons with mean relaxation times $T_1$ up to \SI{407}{\micro\second} at \SI{5}{\giga\hertz}. Our transmons are supplemented with a loss characterization study, in which we separate the dominant loss mechanisms and construct a loss budget that agrees with our $T_1$ measurements. Further characterization may establish rhenium as a leading candidate for maximizing decoherence time.
\end{abstract}

\maketitle

\section{Introduction}

Modern superconducting quantum technologies, notably quantum information processing using superconducting circuits, benefit from long decoherence times \cite{sivak2023,krinner2022,nguyen2019}. Large-scale quantum computation, whose performance remains curbed by decoherence \cite{arute2019,wu2021,zhu2022,googlequantumai2023,googlequantumaiandcollaborators2025,gao2025}, holds the promise of exponential speedup in select complexity classes of algorithms \cite{nielsen2010}. A notorious mechanism for such decoherence is the dissipation of electromagnetic field energy in the materials used to host these technologies \cite{martinis2005}.

The transmon qubit \cite{koch2007,schreier2008}, whose internal quality factor $Q_\mathrm{int}$ is related to relaxation time $T_1$ and angular frequency $\omega$ by $Q_\mathrm{int} = \omega T_1$, is regularly used to probe the dissipative properties of a material platform \cite{chang2013,place2021,bal2024,biznarova2024,bland2025}. Among the diverse range of existing superconducting qubits, the realization of a transmon is comparatively easy and the physics well-understood. Importantly, the transmon merely serves as a starting point and should not preclude the generalization of favorable materials to more complex circuits.

The collection of materials that has been explored in the pursuit of longer decoherence times is, by now, extensive. On-chip devices, which capitalize on techniques established by the micro-electromechanical systems (MEMS) industry, consist of superconducting films lithographically patterned on a dielectric substrate. Although microwave cavities are frequently used for quantum information processing owing to their high $Q_\mathrm{int}$ \cite{reagor2016,chakram2021,milul2023}, the transmon (or other nonlinear circuit component of choice) is conventionally fabricated on-chip due to physical limitations imposed by the Josephson junction.

In assessing the merit of an on-chip material platform, a key consideration is dissipation by the substrate through coupling to the electric field of an electromagnetic mode. This is mitigated by choosing a substrate with a low dielectric loss tangent at microwave frequencies and cryogenic temperatures, such as sapphire \cite{creedon2011,read2023,ganjam2024} or silicon \cite{oconnell2008,checchin2022,zhang2024,bland2025}. The superconductor itself suffers dissipation through coupling to the magnetic field. Desirable properties for the superconductor include a critical temperature much higher than the operating temperature \cite{mattis1958} and a clean microstructure with a low level of disorder \cite{grunhaupt2018,amin2022,gupta2025}. These are not highly restrictive demands, and many superconductors---tantalum, niobium and aluminum, to highlight a few---are deposited in a manner that already meets these requirements.

The next consideration is the interfaces. We are concerned with three types of interfaces: metal-air (MA), metal-substrate (MS), and substrate-air (SA). Dielectric loss in these interfaces, which are often collectively referred to as the ``surface", is identified as a dominant loss mechanism in state-of-the-art transmons \cite{wang2015,crowley2023,ganjam2024,bland2025}. For transmons that do not use aluminum for the electrodes of the shunt capacitor (``pads") but still employ the popular Al/AlO$_x$/Al superconductor-insulator-superconductor junction, there is the additional task of integrating different superconducting films into one circuit with low loss at the interface.

Today, the $T_1$ of a well-performing transmon sits comfortably beyond \SI{100}{\micro\second} and in some cases exceeds \SI{1}{\milli\second} \cite{place2021,wang2022,deng2023,bal2024,ganjam2024,kono2024,biznarova2024,tuokkola2025,bland2025}, an advancement that was catalyzed by the introduction of tantalum for the pads \cite{place2021}. These recent developments may be attributed to a clean MS interface, which is treated with piranha solution prior to deposition, and a low-loss native oxide layer at the MA interface. In the case of tantalum, the native oxide is observed to be less lossy when regrown after application of buffered oxide etch (BOE) \cite{mclellan2023,crowley2023}. Nevertheless, interfaces remain a persistent obstacle to extending the $T_1$ of superconducting qubits, for those based on tantalum as well as on other superconductors.

The search for cleaner interfaces continues. In this work, we explore rhenium on sapphire as a direction for improving qubit performance. Along with tantalum and niobium, rhenium is a superconducting refractory metal, exhibiting a bulk critical temperature of \SI{1.7}{\kelvin} \cite{hulm1957}. Characterization of superconducting rhenium films on sapphire has been the subject of prior work \cite{haq1982,song2009,wang2009,sage2011,dumur2016,ratter2017,teknowijoyo2023,ganjam2024a,tarkaeva2025}, where excellent lattice matching between rhenium and sapphire shows promise for a low-loss MS interface \cite{ratter2017}. Remarkably, sputtered rhenium has the capacity to suppress native oxide formation under ambient conditions \cite{ganjam2024a}, which may address the microwave quality of the MA interface, if the native oxide were indeed the source of this loss. Hence there is a strong incentive to investigate this material stack.

In Sec.~\ref{section:fabrication}, we report the $T_1$ of transmons composed of rhenium pads and aluminum junctions on sapphire, and find that they are competitive with, but do not exceed, those of tantalum-based transmons. To understand the origin of losses in our rhenium-based transmons, we analyze, in Sec.~\ref{section:loss}, supplementary loss characterization devices and construct a loss budget for our transmons. This predicted budget is in agreement with our $T_1$ measurements, suggesting that we have obtained a faithful decomposition of loss contributions. We also probe the loss at the rhenium-aluminum interface and confirm its contribution to be negligible, allowing rhenium to be incorporated into conventional aluminum processes without sacrificing $Q_\mathrm{int}$. Our comparison between rhenium and tantalum in Sec.~\ref{section:discussion} has implications for the impact of the native oxide (or lack thereof) on $Q_\mathrm{int}$.

\section{Fabrication \& measurement}\label{section:fabrication}

\begin{figure*}[t]
    \includegraphics{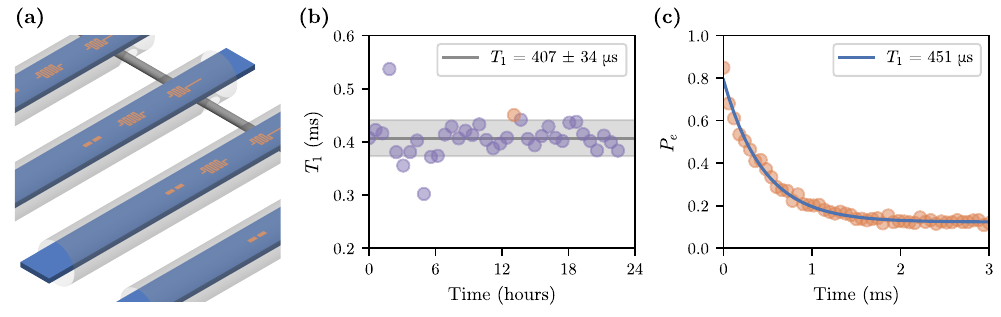}
    \caption{\label{fig:T1} \textbf{Measurement of rhenium-based transmons.} \textbf{(a)}~Schematic of our transmon chips, measured in hanger configuration in a superconducting tunnel package. Each chip consists of a transmon dispersively coupled to a stripline-based readout resonator and Purcell filter. \textbf{(b)}~Relaxation time $T_1$ of our best-performing transmon (``Transmon 5") over the duration of 24 hours. Each data point (purple) is an averaged $T_1$ trace, along with the temporal mean and standard deviation underlaid in gray. \textbf{(c)}~Averaged $T_1$ trace (orange) for the highlighted data point in \textbf{(b)}, fit by an exponential decay (blue).}
\end{figure*}

We fabricate and measure five transmons in the same experimental setup as in Ref.~\cite{ganjam2024} at an operating temperature of \SI{20}{\milli\kelvin}. As depicted in Fig.~\ref{fig:T1}(a), each transmon is dispersively coupled to a stripline-based readout resonator and Purcell filter, measured in hanger configuration in a tunnel package machined from 6061 aluminum alloy. Control and readout of these ``3D transmons", as they are known colloquially when the package is a microwave cavity that serves as electrical ground \cite{paik2011}, are enabled by capacitively coupling to a coaxial transmission line. The \SI{150}{\nano\meter}-thick rhenium film is deposited by sputtering at \SI{900}{\celsius} and patterned by reactive-ion etching (RIE). Sapphire grown using the heat-exchanger method (HEM) and annealed at \SI{1200}{\celsius} is chosen as the substrate for its low dielectric loss tangent \cite{read2023,ganjam2024}. The \SI{50}{\nano\meter}-thick Al/AlO$_x$/Al film that contains the junction is deposited by double-angle electron-beam evaporation and patterned by lift-off using the conventional Dolan bridge technique \cite{dolan1977}, yielding approximately \SI{9}{\nano\henry} for the inductance of the junction $L_\mathrm{J}$. The complete fabrication process is detailed in Appendix~\ref{appendix:fabrication}.

Table~\ref{tab:T1} lists the measured frequency and $T_1$ of the transmons. $T_1$ is sampled over a duration on the order of 24 hours to obtain a temporal mean and standard deviation. Figure~\ref{fig:T1}(b) shows the temporal behavior of our best-performing transmon (``Transmon 5"), while Fig.~\ref{fig:T1}(c) shows an individual $T_1$ trace, fit by an exponential decay. We report a sample average $Q_\mathrm{int}$ of \num{9.1(2.5)e6}, which corresponds to a $T_1$ of \SI{289(80)}{\micro\second} at \SI{5}{\giga\hertz}. Significant device-to-device variation in $T_1$ is observed, as well as temporal fluctuation, both of which are widely reported in literature and modeled by the coupling of the electric field to two-level systems (TLS) in the dielectric environment \cite{muller2015,klimov2018,burnett2019,schlor2019,bejanin2021,carroll2022}. As we will discuss in the following sections, a major contributor may be the variation in the quality of the aluminum near-junction region, where, for its footprint, the surface participation ratio is disproportionately high.

\begin{table}[h]
    \begin{ruledtabular}
    \begin{tabular}{
        c
        S[table-format=1.2]
        S[table-format=3(2)]
        S[table-format=2.1(1.1)]
    }
         & {$\omega/2\pi$ (\si{\giga\hertz})} & {$T_1$ (\si{\micro\second})} & {$Q_\mathrm{int}$ (\num{e6})} \\
         \midrule
         Transmon 1 & 5.00 & 297(22) &  9.3( 7) \\
         Transmon 2 & 5.45 & 277(18) &  9.5( 6) \\
         Transmon 3 & 5.48 & 201(17) &  6.9( 6) \\
         Transmon 4 & 5.26 & 205(40) &  6.8(13) \\
         Transmon 5 & 5.06 & 407(34) & 13.0(11) \\
    \end{tabular}
    \end{ruledtabular}
    \caption{\label{tab:T1} \textbf{Relaxation times of transmons.} Relaxation time $T_1$ is sampled over a duration on the order of 24 hours. The mean and standard deviation capture temporal fluctuation over this timescale. Internal quality factor $Q_\mathrm{int}$ is related to $T_1$ and angular frequency $\omega$ by $Q_\mathrm{int}=\omega T_1$.}
\end{table}

\section{Loss characterization}\label{section:loss}

\begin{figure*}[t]
    \includegraphics{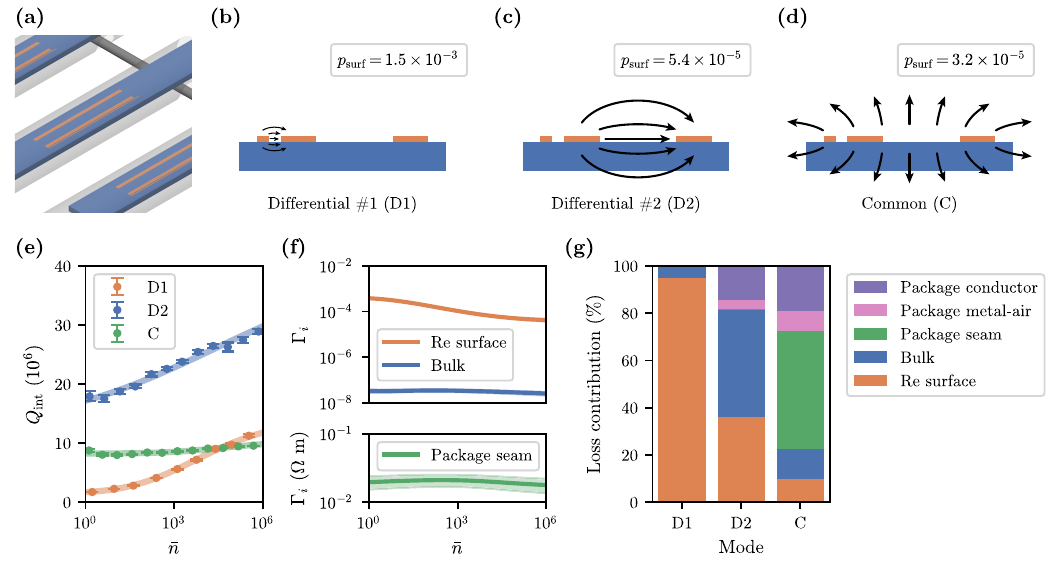}    \caption{\label{fig:tripole} \textbf{Loss analysis of the tripole stripline resonator.} \textbf{(a)}~Schematic of our tripole stripline resonators, measured in hanger configuration in a superconducting tunnel package. The \SI{10}{\micro\meter}-wide strip is not visible at this scale. \textbf{(b)}--\textbf{(d)}~Cross-sectional illustrations of the electric field profile of the first differential (D1), second differential (D2) and common (C) modes of the tripole stripline resonator respectively. The strip widths and separations are not to scale. The D1 mode has the highest surface participation $p_\mathrm{surf}$, rendering it more sensitive to dielectric loss at the interfaces. \textbf{(e)}~Dependence of the internal quality factor $Q_\mathrm{int}$ on the circulating photon number $\bar{n}$ for each of the three fundamental modes of a representative tripole stripline resonator (``Tripole~1"). The data points are fit by the two-level system (TLS) model, allowing extraction of $Q_\mathrm{int}$ at single-photon power, $\bar{n}=1$. \textbf{(f)}~Power dependence of the loss factors $\Gamma_i$ for Tripole~1, extracted using $Q_\mathrm{int}$ data from \textbf{(e)} and inverting the participation matrix. Unlike surface and bulk, the package seam loss factor is not dimensionless and has been plotted separately. The shaded area represents the error propagated from $Q_\mathrm{int}$ to $\Gamma_i$, and is too small to be visible for surface and bulk. \textbf{(g)}~Relative loss budget at $\bar{n}=1$ for the three fundamental modes of Tripole~1, constructed using loss factors from \textbf{(f)} and participations from simulation. The contribution by package seam is too small to be visible in the budgets for the D1 and D2 modes; similarly for package metal-air and package conductor in the D1 mode.}
\end{figure*}

\begin{figure*}[t]
    \includegraphics{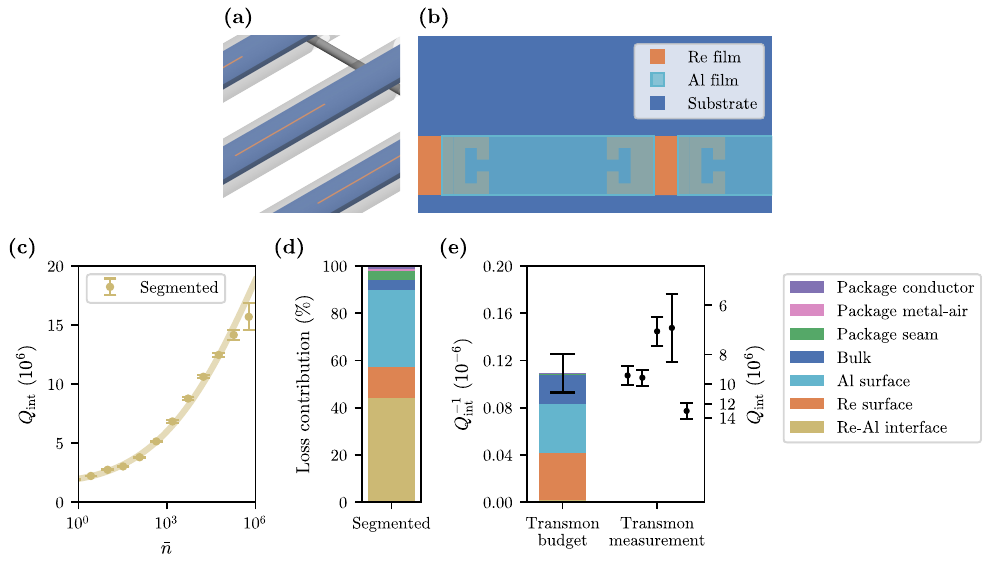}
    \caption{\label{fig:transmon} \textbf{Constructing the loss budget for the transmon.} \textbf{(a)}~Schematic of our segmented stripline resonators, measured in hanger configuration in a superconducting tunnel package. \textbf{(b)}~Top-down illustration of the alternating rhenium and aluminum segments in the strip. The notch geometry replicates the rhenium-aluminum interface in the transmon. \textbf{(c)}~Dependence of the internal quality factor $Q_\mathrm{int}$ on the circulating photon number $\bar{n}$ for the fundamental mode of a representative segmented stripline resonator (``Segmented 1"). The data points are fit by the two-level system (TLS) model, allowing extraction of $Q_\mathrm{int}$ at single-photon power, $\bar{n}=1$. \textbf{(d)}~Relative loss budget at $\bar{n}=1$ for the fundamental mode of Segmented 1, constructed using $Q_\mathrm{int}$ data from \textbf{(c)} and participations from simulation. \textbf{(e)}~Absolute loss budget of the transmon in comparison to the $Q_\mathrm{int}$ of measured transmons in Table~\ref{tab:T1}. The error in the loss budget is propagated from the sample standard deviation of $\Gamma_i$. The contributions by the rhenium-aluminum interface and the package are too small to be visible in the budget.}
\end{figure*}

Our loss characterization study follows the participation matrix approach established by existing works \cite{woods2019,melville2020,lei2023,ganjam2024}. Within this framework, linear resonant modes are used as a proxy for investigating loss in qubits. We write the $Q_\mathrm{int}$ of such a mode as a sum of loss contributions,
\begin{equation}
    Q_\mathrm{int}^{-1} = \sum_i p_i\Gamma_i.
    \label{eq:Qint}
\end{equation}
The two quantities into which each term is decomposed are the participation ratio (``participation") $p_i$, the fraction of electromagnetic field energy stored in a lossy region, and the loss factor $\Gamma_i$, which quantifies how lossy that region is. Conceptually, the loss factor is the generalization of the dielectric loss tangent to all types of losses. The motivation behind these definitions is to contain the geometric dependence of the loss contribution within the participation.

The objective of our loss characterization study is to extract $\Gamma_i$, which is transferable to other modes engineered through the same fabrication process. While we can compute $p_i$ in electromagnetic simulation and measure $Q_\mathrm{int}$ experimentally, in the case where there is more than one term in the sum, we do not have enough equations to solve for all values of $\Gamma_i$ using only one mode. Rather, using $n$ modes, we can solve for $n$ values of $\Gamma_i$. This is equivalent to, mathematically, solving a system of $n$ simultaneous equations. We write our system of equations in a matrix formalism as
\begin{equation}
    \mathbf{K} = \mathbf{P}\mathbf{\Gamma},
    \label{eq:matrix}
\end{equation}
where $\mathbf{K}$ is a vector of $\kappa_i=Q_{\mathrm{int},i}^{-1}$ for the $i^\mathrm{th}$ mode, $\mathbf{\Gamma}$ is a vector of $\Gamma_j$ for the $j^\mathrm{th}$ loss mechanism, and they are related by $\mathbf{P}$, the matrix of participations $p_{ij}$. Solving for $\mathbf{\Gamma}$ requires inversion of $\mathbf{P}$.

In practice, $\mathbf{K}$ receives measurement error. To mitigate error propagation from $\mathbf{K}$ to $\mathbf{\Gamma}$, we want the matrix to be well-conditioned; the ideal case is for $\mathbf{P}$ to be a diagonal matrix. This translates to the problem of designing each mode to be sensitive to a different loss mechanism. Ultimately, any mode will receive contributions from multiple loss mechanisms, but by amplifying the sensitivity to a particular loss mechanism, we minimize the error propagated from $\mathbf{K}$ to $\mathbf{\Gamma}$.

In our device architecture, we identify the following loss mechanisms as potentially significant. Firstly, as noted earlier, the three interfaces associated with the superconducting rhenium film---MA, MS and SA---give rise to dielectric loss. These are grouped into the ``surface", where the corresponding surface loss factor $\Gamma_\mathrm{surf}$ is in reality an average of the component loss factors, weighted by their participations. Secondly, there exists dielectric loss associated with the bulk of the substrate, distinct from the surface. Lastly, we identify losses associated with the superconducting aluminum package: dielectric loss at the MA interface of the package, conductor loss in the package, and loss at the seam where the separately machined components of the package meet. Mathematical definitions of $p_i$ and $\Gamma_i$ for these loss mechanisms are given in Appendix~\ref{appendix:simulation}.

To separate surface, bulk and package seam, we use the tripole stripline resonator, measured in the same experimental setup as the transmons, as depicted in Fig.~\ref{fig:tripole}(a). A detailed description of the tripole stripline resonator is introduced in Ref.~\cite{ganjam2024}. To summarize, the resonator has three fundamental modes: the narrow-gap first differential (D1) mode, sensitive to the surface; the wide-gap second differential (D2) mode, sensitive to the bulk; the common mode, sensitive to the package. The electric field profiles of these three modes are illustrated in Fig.~\ref{fig:tripole}(b)--\ref{fig:tripole}(d) respectively. We choose to focus on the $\lambda/2$ modes, although the higher-order modes bear similar sensitivities and would in principle work equally as effectively. Our tripole stripline resonators consist of a rhenium film on a sapphire substrate, and are fabricated on the same wafer in the same process as our transmons. We account for package MA and package conductor using loss factors extracted by the forky whispering-gallery-mode resonator (FWGMR) and the ellipsoidal cavity in a prior work \cite{lei2023}, which are machined from the same material as our packages. The values of $\Gamma_i$ that we transfer to our work are given in Appendix~\ref{appendix:analysis}.

Figures~\ref{fig:tripole}(e)--\ref{fig:tripole}(g) present our procedure for analyzing the tripole stripline resonator. $Q_\mathrm{int}$ is extracted through measurement of the scattering matrix element $S_{21}$ using a vector network analyzer (VNA), followed by the conventional circle fitting procedure \cite{gao2008,khalil2012,probst2015}. Frequency is sampled uniformly around the circle for efficiency \cite{baity2024,ganjam2024} and modes are generally undercoupled to minimize uncertainty due to Fano interference \cite{rieger2023}. For each mode, $Q_\mathrm{int}$ is measured as a function of the circulating photon number $\bar{n}$. For a mode measured in hanger configuration, $\bar{n}$ is related to the input power $P$ by
\begin{equation}
    \bar{n}=\frac{2}{\hbar\omega^2}\frac{Q_\mathrm{tot}^2}{Q_\mathrm{ext}}P,
    \label{eq:nbar}
\end{equation}
where $Q_\mathrm{ext}$ and $Q_\mathrm{tot}$ are the external (coupling) and total (loaded) quality factors respectively \cite{ganjam2024}.

We assert that the power dependence of $Q_\mathrm{int}$ is dielectric in origin and is modeled by the coupling of the electric field to a bath of TLSs in dielectric materials \cite{gao2008,pappas2011}, described by
\begin{equation}
    Q_\mathrm{int}^{-1}\left(\bar{n}\right)= Q_0^{-1}+\frac{Q_1^{-1}}{\sqrt{1+\left(\frac{\bar{n}}{n_\mathrm{c}}\right)^\beta}}.
    \label{eq:TLS}
\end{equation}
This model supposes an asymptotic high-power $Q_\mathrm{int}^{-1}=Q_0^{-1}$, an asymptotic low-power $Q_\mathrm{int}^{-1}=Q_0^{-1}+Q_1^{-1}$, and a smooth transition characterized by the critical photon number $n_\mathrm{c}$ and a parameter $\beta$ of order unity that determines the width of this transition. Phenomenologically, $Q_1^{-1}$ captures all power-dependent loss and $Q_0^{-1}$ all power-independent loss in the power regime of interest.

Fitting Eq.~(\ref{eq:TLS}) to our power sweep of $Q_\mathrm{int}$ allows interpolation of $Q_\mathrm{int}$ over a continuous range of $\bar{n}$. By inverting the participation matrix $\mathbf{P}$, we solve our system of simultaneous equations and draw out the dependence of $\Gamma_i$ on $\bar{n}$. As shown in Fig.~\ref{fig:tripole}(f), the values of $\Gamma_i$ are extracted with reasonable error, a consequence of our sufficiently well-conditioned $\mathbf{P}$. Upon comparing the loss factors for the surface $\Gamma_\mathrm{surf}$ and for the sapphire bulk $\Gamma_\mathrm{bulk}$, we see that, as is widely reported in literature for a broad variety of interfaces, $\Gamma_\mathrm{surf}$ is orders of magnitude higher \cite{creedon2011,woods2019,melville2020,read2023,crowley2023,ganjam2024,bland2025}. Notably, $\Gamma_\mathrm{surf}$ is more power-dependent, varying by an order of magnitude over the measured range of $\bar{n}$, which is ascribed to the higher density of TLSs present in the dielectric interfaces.

We are interested in the breakdown at single-photon power, $\bar{n}=1$, the regime in which a transmon operates for quantum information processing. In Fig.~\ref{fig:tripole}(g), we construct the loss budget for each mode of the tripole stripline resonator, where the contribution by the $i^\mathrm{th}$ loss mechanism is given by $p_i\Gamma_i$. Its reciprocal $\left(p_i\Gamma_i\right)^{-1}$ may be interpreted as the limit on $Q_\mathrm{int}$ imposed by that loss mechanism. By design, the composition of each budget varies significantly. The D1 mode receives an overwhelmingly large contribution from surface loss, the D2 mode receives an elevated contribution from bulk loss, and the common mode an elevated contribution from package losses. The varying sensitivity of each mode to different loss mechanisms, which has enabled the separation of those loss mechanisms, is confirmed by these budgets.

Several assumptions are implicit in our model of loss: (1)~The region associated with each loss mechanism is assumed to be homogeneous, which may not be true if, for instance, the bulk of the substrate exhibits a spatially varying loss tangent. Such significant inhomogeneity has not been observed in our sapphire substrates \cite{read2023,ganjam2024}. (2)~Each loss mechanism is assumed to be frequency-independent over the frequency range of interest (\SIrange{4}{8}{\giga\hertz}), which is true for the TLS model of dielectric loss in the case of $\hbar\omega \gg k_\mathrm{B}T$. (3)~The loss factors are assumed to be transferable from device to device. In our case, the package loss factors extracted by the FWGMR and ellipsoidal cavity are assumed to be the same as for our packages. (4)~The quantity $\Gamma_\mathrm{surf}$, derived by combining the loss factors from the three types of interfaces, is only applicable to geometries where the participations of the three interfaces are present in the same proportions. This is largely true in a planar architecture. (5)~We neglect, naturally, any loss mechanisms that are excluded from our model, such as conductor loss in the film. The power-independent component of $Q_\mathrm{int}$ can place an approximate bound on the contribution by conductor loss, which is negligibly small for our devices.

Our transmons with rhenium pads and aluminum junctions also receive contributions from the interfaces at the aluminum film, and from the interface between the rhenium and aluminum films. We include the contribution by aluminum surface using loss factors extracted by aluminum tripole stripline resonators in a prior work \cite{ganjam2024}, fabricated using the same process as that for the junction.

Loss at rhenium-aluminum interface is isolated using the segmented stripline resonator, a device whose strip consists of alternating rhenium and aluminum segments, fabricated on the same wafer as our transmons and measured in the same package, as depicted in Fig.~\ref{fig:transmon}(a). A detailed description of the segmented stripline resonator is introduced in Ref.~\cite{ganjam2024}. The motivation behind the design of this device is to substantially increase the number of rhenium-aluminum interfaces beyond that in the transmon to magnify sensitivity to loss at that interface. An illustration of the segments is provided in Fig.~\ref{fig:transmon}(b).

Figures~\ref{fig:transmon}(c),~\ref{fig:transmon}(d) present our procedure for analyzing the segmented stripline resonator, in the same manner as that for the tripole stripline resonator. The remaining loss mechanisms are accounted for by transferring loss factors from separate devices, including the rhenium tripole stripline resonators in this work.

We are now in a position to construct the loss budget for our transmons. As we have done for the modes of the tripole stripline resonator, the contribution by the $i^\mathrm{th}$ loss mechanism is given by $p_i\Gamma_i$, where this time the participations have been computed for our transmon design. Our loss budget, shown in Fig.~\ref{fig:transmon}(e), predicts a $Q_\mathrm{int}$ of \num{9.2(1.4)e6}, which corresponds to a $T_1$ of \SI{292(44)}{\micro\second} at \SI{5}{\giga\hertz} and is in agreement with our measured sample average $Q_\mathrm{int}$ of \num{9.1(2.5)e6}. Importantly, the budget reveals critical regions to address for future investigation, which, in our case, are the surfaces. Although the aluminum near-junction region occupies an area orders of magnitude smaller than that of the rhenium pads, the corresponding surface participations, tabulated in Appendix~\ref{appendix:analysis}, do not scale accordingly, a consequence of concentrating the mode's inductance into an element as compact as a Josephson junction. This suggests that the detrimental effect of TLSs in the near-junction region is amplified, leading to larger $Q_\mathrm{int}$ fluctuation than for a more dilute mode with a comparable $Q_\mathrm{int}$.

\section{Discussion}\label{section:discussion}

Table~\ref{tab:contributions} lists the sample average loss factors that we have extracted at $\bar{n}=1$ using our tripole and segmented stripline resonators, as well as loss factors that we transfer from prior works for the purpose of constructing the loss budget for our transmons. These values are more insightful when compared with other processes, so we choose to compare with loss factors from our tantalum on sapphire process \cite{ganjam2024} in Table~\ref{tab:tantalum}. It is worth noting that the precise value of $\Gamma_\mathrm{surf}$ is dependent on the surface participation $p_\mathrm{surf}$, which in turn is strongly influenced by the manner in which the simulation is carried out. Further insight into the computation of participations and their impact on the corresponding loss factors is provided in Appendix~\ref{appendix:simulation}.

Perhaps surprisingly, the rhenium and tantalum surfaces perform similarly, despite the absence of a native oxide for our rhenium films, which we confirm via transmission electron microscopy (TEM) and energy-dispersive X-ray spectroscopy (EDS) in Appendix~\ref{appendix:materials}. This suggests two interpretations. Firstly, the contribution by the MA interface may be minimal in a planar architecture relative to those of MS and SA. This can be elucidated by a loss characterization study that employs a flip-chip architecture, where the MA contribution is isolated by taking advantage of the additional geometric degree of freedom \cite{ganjam2024a}. Secondly, the mechanism by which the MA interface is lossy may not be correlated with oxide thickness, but rather contamination on the surface, likely organic in nature. Unlike tantalum, our rhenium films cannot survive piranha solution; instead, we treat our wafer with concentrated sulfuric acid at \SI{100}{\celsius} after patterning the rhenium film. As sulfuric acid is not as effective an organic cleaning agent as piranha solution, the wafer may not be adequately cleaned. Experimentation with alternate cleaning agents may improve the surface loss factor.

Our loss factor for the rhenium-aluminum interface is comparable to that for the low-loss tantalum-aluminum interface, demonstrating that our ion milling process prior to deposition of the aluminum junctions is effective at preparing the rhenium surface. It is possible that such an aggressive ion milling process is unnecessary given the absence of a native oxide, although this is not tested in this work. Ultimately, any realistic circuit that integrates rhenium and aluminum films bears a similarly high $Q_\mathrm{int}$ limit to that observed in our transmons.

Loss factors that are intuitively expected to remain unaffected by the choice of superconductor and subsequent surface treatments should remain in statistical agreement within our methodology. This is observed for our measurement of annealed HEM sapphire, which imposes a bulk limit on $Q_\mathrm{int}$ of \num{40e6}, for a half-filled dielectric environment with a relative permittivity of 10 (bulk participation of 0.91). Our bulk measurement also approaches that obtained for HEMEX sapphire, the highest grade of HEM sapphire as screened by the manufacturer, in Ref.~\cite{read2023}.

Broadly, we have shown that whichever property that imparts tantalum-based devices high $Q_\mathrm{int}$ is not unique to tantalum, hinting at the importance of chemical treatments more so than the composition of the material itself, at present. Chemical profiling of the MA interface, as tantalum has received \cite{mclellan2023}, and further experimentation with the deposition process and chemical treatments may enable even higher decoherence times with rhenium.

\begin{table*}
    \begin{ruledtabular}
    \begin{tabular}{
        c
        S[table-format=1.1(1.1)e-2]
        c
        S[table-align-comparator=false,table-format=<2]
        S[table-align-comparator=false,table-format=>4]
    }
        &
        &
        & \multicolumn{2}{c}{Loss contribution} \\
        \cmidrule{4-5}
        Loss
        & {Loss factor, $\Gamma_i$}
        & Reference
        & {Relative $p_i\Gamma_i$ (\%)}
        & {$\left(p_i\Gamma_i\right)^{-1}$ (\num{e6})} \\
        \midrule
        Re-Al interface (\si{\ohm\meter}) & 2.9(8)e-12 & This work & 2 & 486 \\
        Re surface & 3.6(6)e-4 & This work & 36 & 25 \\
        Al surface & 8.7(24)e-4 & \cite{ganjam2024} & 38 & 24 \\
        Bulk & 2.9(11)e-8 & This work & 23 & 40 \\
        Package seam (\si{\ohm\meter}) & 1.1(8)e-2 & This work & <1 & >1000 \\
        Package metal-air & 8.4(47)e-2 & \cite{lei2023} & <1 & >1000 \\
        Package conductor (\si{\ohm}) & 5.3(17)e-6 & \cite{lei2023} & <1 & >1000 \\
    \end{tabular}
    \end{ruledtabular}
    \caption{\label{tab:contributions} \textbf{Loss contributions to the transmon.} The product of the participation $p_i$ and loss factor $\Gamma_i$ gives the absolute contribution by the $i^\mathrm{th}$ loss mechanism, while its reciprocal $\left(p_i\Gamma_i\right)^{-1}$ is the limit on $Q_\mathrm{int}$ imposed by that loss. The dominant loss contributions to our rhenium-based transmons are the surfaces, followed by the bulk. Rhenium surface, bulk and package seam are separated using tripole stripline resonators in this work. Rhenium-aluminum interface is isolated using segmented stripline resonators in this work. Aluminum surface is transferred from tripole stripline resonators in Ref.~\cite{ganjam2024}. Package metal-air and package conductor are transferred from forky whispering-gallery-mode resonators and ellipsoidal cavities in Ref.~\cite{lei2023}.}
\end{table*}

\begin{table}
    \begin{ruledtabular}
    \begin{tabular}{
        c
        S[table-format=1.1(1.1)]
        S[table-format=1.1(1.1)]
    }
        & \multicolumn{2}{c}{Loss factor, $\Gamma_i$} \\
        \cmidrule{2-3}
        Loss
        & {Rhenium}
        & {Tantalum \cite{ganjam2024}} \\
        \midrule
        X-Al interface (\SI{e-12}{\ohm\meter}) & 2.9(8) & 2.6(5) \\
        X surface (\num{e-4}) & 3.6(6) & 3.4(12) \\
        Bulk (\num{e-8}) & 2.9(11) & 4.3(1.9) \\
    \end{tabular}
    \end{ruledtabular}
    \caption{\label{tab:tantalum} \textbf{Comparison of rhenium and tantalum processes.} Loss factors $\Gamma_i$ for our rhenium on sapphire process are extracted by this work, and compared with the tantalum on sapphire process from Ref. \cite{ganjam2024}. X denotes the superconductor of interest (Re, Ta). The tantalum surface loss factor has been corrected for participation convergence, as explained in Appendix~\ref{appendix:simulation}.}
\end{table}

\begin{acknowledgments}
We thank Heekun Nho and John Garmon for assistance in implementing our transmon measurements. We thank Alex Read for assistance in conceiving our mesh convergence study. We thank Nico Zani and Archan Banerjee for helpful discussions. Fabrication and characterization facilities were partially supported by the Yale University Cleanroom, the Yale Institute for Nanoscience and Quantum Engineering (YINQE) and the Yale University West Campus Materials Characterization Core (MCC). We thank Yong Sun, Lauren McCabe, Yeongjae Shin, Kelly Woods and Michael Rooks for assistance in implementing our fabrication process. We thank Tyler Wang, Shize Yang, Sungwoo Sohn and Min Li for assistance in materials characterization. Computing resources were partially supported by the Yale Center for Research Computing (YCRC). We thank Misha Guy for assistance in implementing our simulations on the high-performance computing cluster. We acknowledge the support of the Yale Quantum Institute (YQI).

This research was sponsored by the Army Research Office (ARO) under grant no. W911NF-23-1-0051, and by the U.S. Department of Energy (DoE), Office of Science, National Quantum Information Science Research Centers, and Co-design Center for Quantum Advantage (C2QA) under contract number DE-SC0012704. The views and conclusions contained in this document are those of the authors and should not be interpreted as representing the official policies, either expressed or implied, of the ARO, DoE or the U.S. Government. The U.S. Government is authorized to reproduce and distribute reprints for Governmental purposes notwithstanding any copyright notation herein.

L.F. and R.J.S. are consultants and shareholders of D-Wave Quantum Inc.
\end{acknowledgments}

\section*{Author contributions}

S.G. conceptualized the work. S.G. and Y.W. developed the fabrication process. Y.W. fabricated and measured the devices, and performed the analysis and simulations. I.N. performed the materials characterization. L.F. and R.J.S. supervised the work. Y.W. wrote the manuscript with feedback from all authors.

\appendix

\section{FABRICATION}\label{appendix:fabrication}

We use 4-inch, $c$-plane sapphire wafers grown using HEM. The wafer is treated with piranha solution consisting of a 2:1 mixture of concentrated sulfuric acid and 30\% hydrogen peroxide for 20 minutes, followed by a rinse in deionized water for 20 minutes. The wafer is annealed at \SI{1200}{\celsius} for 1 hour in an oxygen-rich environment (FirstNano EasyTube 6000). Our deposition process takes place in the Kurt J. Lesker CMS-18 sputtering system, which reaches a chamber pressure of \SI{e-9}{\torr}. We first carry out a dehydration bake at a substrate temperature of \SI{400}{\celsius} for 15 minutes, followed by deposition of rhenium at \SI{900}{\celsius} for 30 minutes at a deposition rate of approximately \SI{0.8}{\angstrom\per\second} with an argon pressure of \SI{2}{\milli\torr}. The wafer is allowed to thermalize at \SI{900}{\celsius} for 15 minutes prior to deposition.

To pattern the rhenium film, the wafer is primed with hexamethyldisiliazane (HMDS) for 25 minutes in the YES 310TA. Microposit S1827 resist is deposited by spin-coating, patterned in the Heidelberg MLA 150, developed in Microposit MF-319, and hard-baked at \SI{120}{\celsius} for 1 minute. The wafer is treated with oxygen plasma at \SI{300}{\milli\torr} and \SI{150}{\watt} for 2 minutes (Autoglow 200). We carry out reactive-ion etching using SF\textsubscript{6} with a flow rate of \SI{20}{sccm}, a pressure of \SI{10}{\milli\torr}, and an RF power of \SI{50}{\watt} for 13 minutes (Oxford Plasmalab 80 Plus). The resist mask is removed by sonication in \textit{N}-methyl-2-pyrrolidone (NMP), acetone and isopropanol (IPA), followed by a rinse in deionized water.

Before proceeding to junction fabrication, we treat the wafer with the following: concentrated sulfuric acid at \SI{100}{\celsius} for 20 minutes, a rinse in deionized water for 20 minutes, 10:1 BOE for 5 minutes, and a rinse in deionized water again for 20 minutes.

To pattern the aluminum film, we carry out a dehydration bake under rough vacuum at \SI{150}{\celsius} for 25 minutes (YES 310TA). MMA(8.5)MAA EL13 and PMMA A4 bilayer resist is deposited by spin-coating, patterned in the Raith EBPG 5200, and developed in a 3:1 mixture of IPA and deionized water at \SI{6}{\celsius}. Our double-angle deposition process takes place in the Plassys UMS 300 electron-beam evaporation system, which reaches a chamber pressure of \SI{e-10}{\torr}. The process consists of argon ion milling at \SI{\pm45}{\degree} for 20 seconds at each angle, a \SI{20}{\nano\meter} deposition at \SI{-25}{\degree}, an oxidation at \SI{30}{\torr} for 10 minutes, a \SI{30}{\nano\meter} deposition at +\SI{25}{\degree}, and an oxidation at \SI{50}{\torr} for 5 minutes. The lift-off process takes place in NMP at \SI{80}{\celsius} for 1 hour. The wafer is treated with sonication in NMP, acetone and IPA, followed by a rinse in deionized water.

To protect the wafer during the dicing process, we prime the wafer with HMDS (YES 310TA) and deposit Microposit S1827 resist by spin-coating. The wafer is diced in the ADT Provectus 7100. The resist is removed by NMP at \SI{80}{\celsius} for 10 minutes, followed by sonication in NMP, acetone and IPA, and a rinse in deionized water.

\section{MATERIALS CHARACTERIZATION}\label{appendix:materials}

\begin{figure*}
    \includegraphics{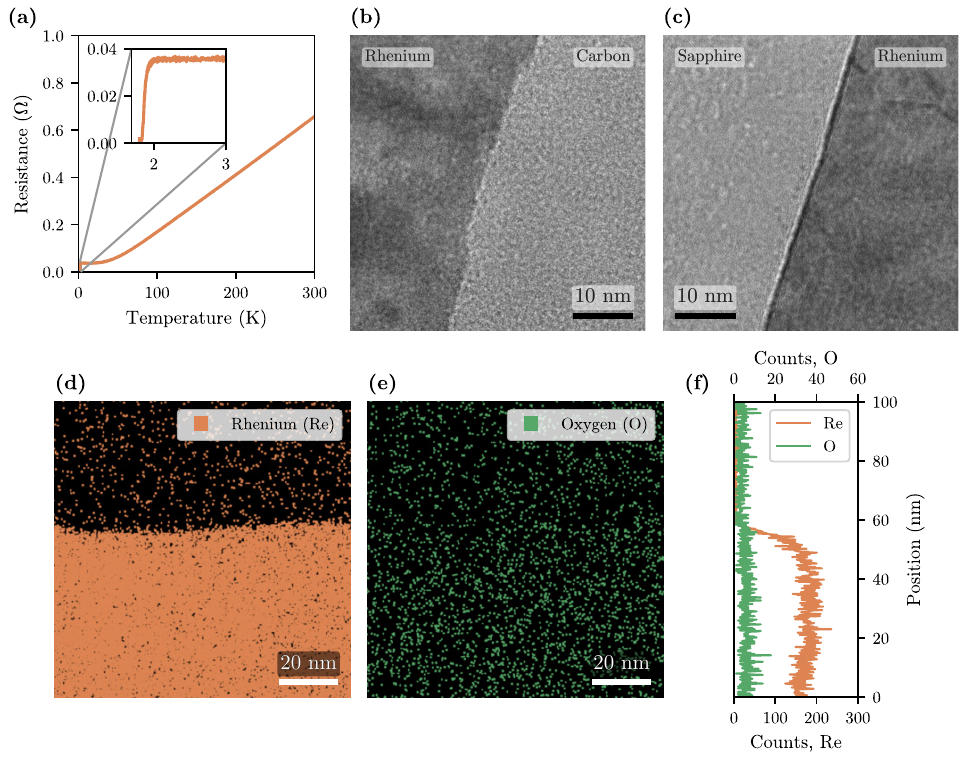}
    \caption{\label{fig:materials} \textbf{Materials characterization of the rhenium film}. \textbf{(a)}~Temperature dependence of electrical resistance, indicating the onset of the superconducting transition at \SI{1.9}{\kelvin} and a residual-resistivity ratio (RRR) of 19. The temperature floor of the measurement system is \SI{1.8}{\kelvin}. \textbf{(b)}~Transmission electron micrograph (TEM) of the metal-air (MA) interface with rhenium on the left and carbon on the right, demonstrating the absence of a native oxide layer on the nanometer scale. The carbon is introduced as part of TEM sample preparation. \textbf{(c)}~TEM of the metal-substrate (MS) interface with sapphire on the left and rhenium on the right. \textbf{(d),~(e)} Distribution of rhenium and oxygen counts respectively at the MA interface, obtained by energy dispersive X-ray spectroscopy (EDS). The orientation and magnification of these data compared to \textbf{(b)} are different. Colored pixels, which are exaggerated for clarity, indicate the location of at least one count. \textbf{(f)}~Depth profile of rhenium and oxygen counts obtained by EDS. The number of counts along the horizontal axis of \textbf{(d),~(e)} is summed to generate this plot. The elements are plotted on different axes to exaggerate the oxygen count, where no elevated oxygen count beyond the background is visible at the MA interface. The transition from metal to air in the rhenium count is not sharp due to the imprecise alignment of the interface along the horizontal axis.}
\end{figure*}

We measure the temperature dependence of the electrical resistance of a \SI{4}{\milli\meter} $\times$ \SI{3}{\milli\meter} rhenium on sapphire sample in the Quantum Design PPMS DynaCool. As shown in Fig.~\ref{fig:materials}(a), we observe the onset of the superconducting transition at approximately \SI{1.9}{\kelvin}, similar to the critical temperature reported for rhenium films deposited on sapphire under similar conditions \cite{haq1982,song2009,dumur2016,ratter2017}. 

To verify the absence of a native oxide, we perform TEM and EDS. Samples are prepared in the ThermoFisher Helios G4 UX DualBeam and imaged in the FEI Tecnai Osiris. In contrast to the distinct \SI{3}{\nano\meter}-thick native oxide on tantalum films observed in TEM in literature \cite{place2021,mclellan2023,ganjam2024}, no such layer is present in Fig.~\ref{fig:materials}(b), a TEM image of the MA interface. We also display, in Fig.~\ref{fig:materials}(c), the MS interface between rhenium and sapphire. The elemental distributions of rhenium and oxygen obtained by EDS reveal no visibly elevated oxygen concentration at the MA interface, as shown in Fig.~\ref{fig:materials}(d)--(f).

\section{DEVICE DESIGN}\label{appendix:design}

\begin{figure*}
    \includegraphics{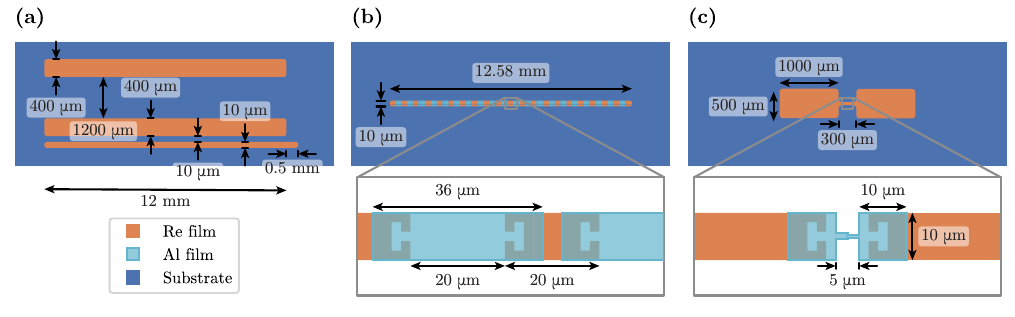}
    \caption{\label{fig:design} \textbf{Designs of devices used in this work.} \textbf{(a)--(c)} Top-down illustration of the important design parameters in the tripole stripline resonator, segmented stripline resonator and transmon respectively, not to scale. The lengths of the tripole stripline and segmented stripline resonators are marginally varied to separate the resonant frequencies and allow multiplexed measurement.}
\end{figure*}

\subsection{Tripole stripline resonator}

Our tripole stripline resonator design consists of three rhenium strips on a sapphire substrate. Important design parameters are displayed in Fig.~\ref{fig:design}(a). The widths of strips 1, 2 and 3 are \SI{400}{\micro\meter}, \SI{400}{\micro\meter} and \SI{10}{\micro\meter} respectively. The lengths of strips 1 and 2 vary from \SI{11.95}{\milli\meter} to \SI{12.10}{\milli\meter} to separate the frequencies of nominally identical devices, while strip 3 receives a relative extension of \SI{0.5}{\milli\meter} to decrease the $Q_\mathrm{ext}$ of the D1 mode, which has a comparatively smaller mode volume. Participations are computed for the \SI{12}{\milli\meter} design, although they are negligibly modified by these variations in strip length. The gaps between strips 1 and 2 and between strips 2 and 3 are \SI{1200}{\micro\meter} and \SI{10}{\micro\meter} respectively. Strips 1, 2 and 3 are filleted with a radius of \SI{50}{\micro\meter}, \SI{50}{\micro\meter} and \SI{4}{\micro\meter} respectively. The substrate has a length of \SI{30}{\milli\meter}, a width of \SI{4}{\milli\meter}, and a thickness of \SI{0.65}{\milli\meter}.

\subsection{Segmented stripline resonator}

Our segmented stripline resonator design consists of a single strip of alternating rhenium and aluminum segments on a sapphire substrate. Important design parameters are displayed in Fig.~\ref{fig:design}(b). Each aluminum segment is \SI{36}{\micro\meter} in length; each rhenium segment is \SI{20}{\micro\meter} in length, and terminates in a notch geometry either side of the segment, except for the segments at the ends of the strip. This notch geometry replicates the rhenium-aluminum interface in our transmon design. The length of the strip varies from \SI{12.46}{\milli\meter} (311 pairs of segments) to \SI{12.58}{\milli\meter} (314 pairs of segments) to separate the frequencies of nominally identical devices, and the width is \SI{10}{\micro\meter}. Participations are computed for the \SI{12.58}{\milli\meter} design, although they are negligibly modified by these variations in number of interfaces. The substrate has a length of \SI{40}{\milli\meter}, a width of \SI{4}{\milli\meter}, and a thickness of \SI{0.65}{\milli\meter}.

\subsection{Transmon}

Our transmon design consists of two pads connected by a single lead, at the center of which sits the Josephson junction. Important design parameters are displayed in Fig.~\ref{fig:design}(c). The pads have dimensions of \SI{1000}{\micro\meter} and \SI{500}{\micro\meter} in the directions parallel and perpendicular to the leads respectively. The lead is \SI{300}{\micro\meter} in length and \SI{10}{\micro\meter} in width. Both the pads and the connection between the pads and the lead are filleted with a radius of \SI{50}{\micro\meter}. The lead terminates at a distance of \SI{4.5}{\micro\meter} from the junction in a notch geometry. This notch geometry allows compatibility with alternate methods of junction fabrication such as the Manhattan style \cite{potts2001}. The aluminum contact region has dimensions of \SI{10}{\micro\meter} by \SI{10}{\micro\meter}, and encompasses the region between \SI{7.5}{\micro\meter} and \SI{12.5}{\micro\meter} away from the junction. The wide junction lead is \SI{1}{\micro\meter} in width, while the narrow junction lead, which determines the overlap area and therefore $L_\mathrm{J}$, is \SI{0.23}{\micro\meter} in width. The frequency of the readout resonator is varied to allow multiplexed readout without modifying the participations of the transmon. The substrate has a length of \SI{40}{\milli\meter}, a width of \SI{4}{\milli\meter}, and a thickness of \SI{0.65}{\milli\meter}.

\section{SIMULATION}\label{appendix:simulation}

Our simulations are carried out using the Ansys Electronics suite on Yale University's high-performance computing (HPC) cluster. Surface participations are computed using the same approach as in Ref.~\cite{wang2015}. Because this approach carries with it several technical complexities, we detail our method of computing participations for the devices reported in this work.

Bulk and package participations are computed in a ``global" high-frequency 3D simulation (Ansys HFSS). Due to the small length scale of the near-junction region, surface participations in the near-junction region require a ``local" electrostatic 3D simulation (Ansys Maxwell 3D), while surface participations outside of the near-junction region are supplemented with an electrostatic 2D simulation (Ansys Maxwell 2D) to help resolve the divergence of the electric field at the edges of conducting objects. The global and 2D simulations are stitched together by the concept of a ``donut" region, while the global and local simulations are stitched together by the electric field profile in a ``stitching" region.

\subsection{Simulation of the resonators}

\begin{figure}
    \includegraphics{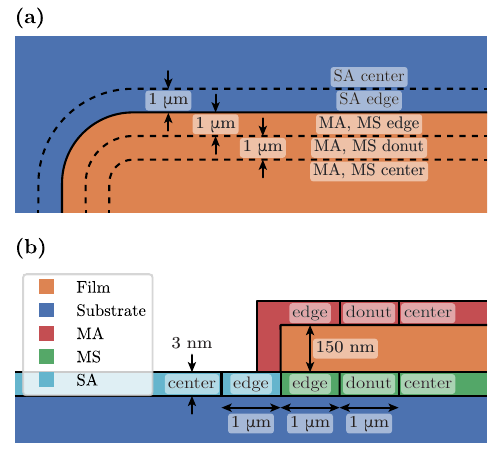}
    \caption{\label{fig:donut} \textbf{Computation of surface participations.} \textbf{(a)}~Top-down illustration of edge, donut and center regions for a generic conducting sheet in the high-frequency 3D simulation (Ansys HFSS). The edge and donut regions are chosen to each have a width of \SI{1}{\micro\meter}. The metal-air (MA) sheet lies \SI{3}{\nano\meter} above the plane of the conducting sheet, while metal-substrate (MS) and substrate-air (SA) lie \SI{3}{\nano\meter} below. \textbf{(b)}~Cross-sectional illustration of edge, donut and center regions in the electrostatic 2D simulation (Ansys Maxwell 2D), with interfaces exaggerated. The ratio of the energy stored in the edge to that in the donut gives the donut-to-edge conversion factor, which is used to derive the energy stored in the edge in the 3D simulation, thereby circumventing the complication of field divergence. The conversion factor for SA uses the MS donut.}
\end{figure}

For the purpose of computing participations, all materials are treated as lossless, under the premise that the field distribution is identical as long as $Q_\mathrm{int}$ is sufficiently high. In the global simulation, the film is modeled as a perfectly conducting sheet and separated into an edge, donut and center region, as illustrated in Fig.~\ref{fig:donut}(a). For all of our designs, the edge and donut regions each have a width of \SI{1}{\micro\meter}, which satisfies the requirement that they be much smaller than the smallest length scale in the design. Donut and center participations are computed in the global simulation, while edge participations are derived from the donut participations via donut-to-edge conversion factors
\begin{equation}
    F = \frac
        {\int_\mathrm{edge} \mathbf{D} \cdot \mathbf{E}~dS}
        {\int_\mathrm{donut} \mathbf{D} \cdot \mathbf{E}~dS},
    \label{eq:factor}
\end{equation}
delivered by the supplemental 2D simulation. Figure~\ref{fig:donut}(b) illustrates how the interfaces are structured around a conducting film in the 2D simulation.

For a design with more than one conducting region of film, such as the tripole stripline resonator, each individual region must be divided into halves. These halves are present in both the global and 2D simulations. This is not required for the segmented stripline resonator, whose 2D simulation consists of a symmetric single film. We choose to fillet vertices in our tripole stripline resonator design, with the expectation that the donut-to-edge conversion factors better capture the field behavior at a filleted vertex than at a square vertex.

In the global simulation, the sapphire substrate is assigned a relative permittivity of 10, and the geometries over which surface participations are computed are drawn as sheets, with the MA sheet lying \SI{3}{\nano\meter} above the conducting sheet and the MS and SA sheets lying \SI{3}{\nano\meter} below the conducting sheet within the substrate. Due to the dielectric environment in which the electric field is evaluated, surface participations are given by
\begin{align}
    p_\mathrm{MA}
    &=\frac{1}{\varepsilon_\mathrm{r}}\frac{t\varepsilon_0\int_\mathrm{MA}\left|\mathbf{E}\right|^2dS}
    {\int_\mathrm{all}\mathbf{D}\cdot\mathbf{E}~dV}, \\
    p_\mathrm{MS}
    &=\varepsilon_\mathrm{r}\frac{t\varepsilon_0\int_\mathrm{MS}\left|\mathbf{E}\right|^2dS}
    {\int_\mathrm{all}\mathbf{D}\cdot\mathbf{E}~dV}, \\
    p_\mathrm{SA}
    &=\varepsilon_\mathrm{r}\frac{t\varepsilon_0\int_\mathrm{SA}\left|\mathbf{E}\right|^2dS}
    {\int_\mathrm{all}\mathbf{D}\cdot\mathbf{E}~dV}.
    \label{eq:surface}
\end{align}
where nominal values of \SI{3}{\nano\meter} and 10 are chosen for the thickness $t$ and the relative permittivity $\varepsilon_\mathrm{r}$ of the interface respectively, a convention adopted by Ref.~\cite{wenner2011,woods2019,melville2020,ganjam2024,gupta2025}.
Note the differing position of $\varepsilon_\mathrm{r}$ in $p_\mathrm{MA}$, a consequence of integrating $\mathbf{E}$ in vacuum in the simulation, whereas in reality the interface is a dielectric. As described previously, surface participations for the edge region are equal to those for the donut region but scaled by the donut-to-edge conversion factor.

The surface participation $p_\mathrm{surf}$ used for the purpose of our loss analysis is the sum of $p_\mathrm{MA}$, $p_\mathrm{MS}$ and $p_\mathrm{SA}$. The corresponding loss factor $\Gamma_\mathrm{surf}$ is the average dielectric loss tangent of these three interfaces, weighted by their participations.

The bulk participation is given by
\begin{equation}
    p_\mathrm{bulk}=\frac{\int_\mathrm{bulk}\mathbf{D}\cdot\mathbf{E}~dV}{\int_\mathrm{all}\mathbf{D}\cdot\mathbf{E}~dV},
    \label{eq:bulk}
\end{equation}
where the corresponding loss factor $\Gamma_\mathrm{bulk}$ is the dielectric loss tangent of the substrate.

The package conductor participation is defined in terms of the ``geometric factor" $G$ by
\begin{equation}
    p_\mathrm{cond}=\frac{1}{G}
    =\frac{\int_\mathrm{cond}\mathbf{B}\cdot\mathbf{H}~dS}{\mu_0\omega\int_\mathrm{all}\mathbf{B}\cdot\mathbf{H}~dV},
    \label{eq:conductor}
\end{equation}
where the corresponding loss factor $\Gamma_\mathrm{cond}$ is the surface resistance. $\Gamma_\mathrm{cond}$ has units of \SI{}{\ohm} and so $p_\mathrm{cond}$ has units of \SI{}{\per\ohm}.

Seam participations are frequently quoted as an admittance per unit length $y_\mathrm{seam}$ \cite{brecht2015,lei2020,lei2023,ganjam2024,krayzman2024}, given by
\begin{equation}
    p_\mathrm{seam}=y_\mathrm{seam}
    =\frac{\int_\mathrm{seam}\left|\mathbf{H}\cdot\hat{\bm{l}}\right|^2~dl}{\omega\int_\mathrm{all}\mathbf{B}\cdot\mathbf{H}~dV},
    \label{eq:seam}
\end{equation}
where the corresponding loss factor $\Gamma_\mathrm{seam}=g_\mathrm{seam}^{-1}$ is the reciprocal of the conductance per unit length. $y_\mathrm{seam}$ and $g_\mathrm{seam}$ have units of \SI{}{\siemens\per\meter}.

Loss at the rhenium-aluminum interface is decomposed into the same quantities as for seam loss, however the participation is computed analytically. For the segmented stripline resonator, the participation is
\begin{equation}
    p_\mathrm{seam}=y_\mathrm{seam}=\frac{2}{\pi Zw}\sum_i \sin^2{\frac{\pi z_i}{l}},
\end{equation}
where $Z$ is the characteristic impedance of the resonator, $w$ is the width of the strip, $l$ is the length of the strip, and the sum is indexed over the $i^\mathrm{th}$ seam at position $z_i$ \cite{ganjam2024}. For the transmon, the participation is
\begin{equation}
    p_\mathrm{seam}=y_\mathrm{seam}=\frac{2}{Zw},
\end{equation}
where $Z$ is the lumped-element impedance of the transmon and $w$ is the width of the lead \cite{ganjam2024}.

\subsection{Simulation of the transmon}

\begin{figure*}
    \includegraphics{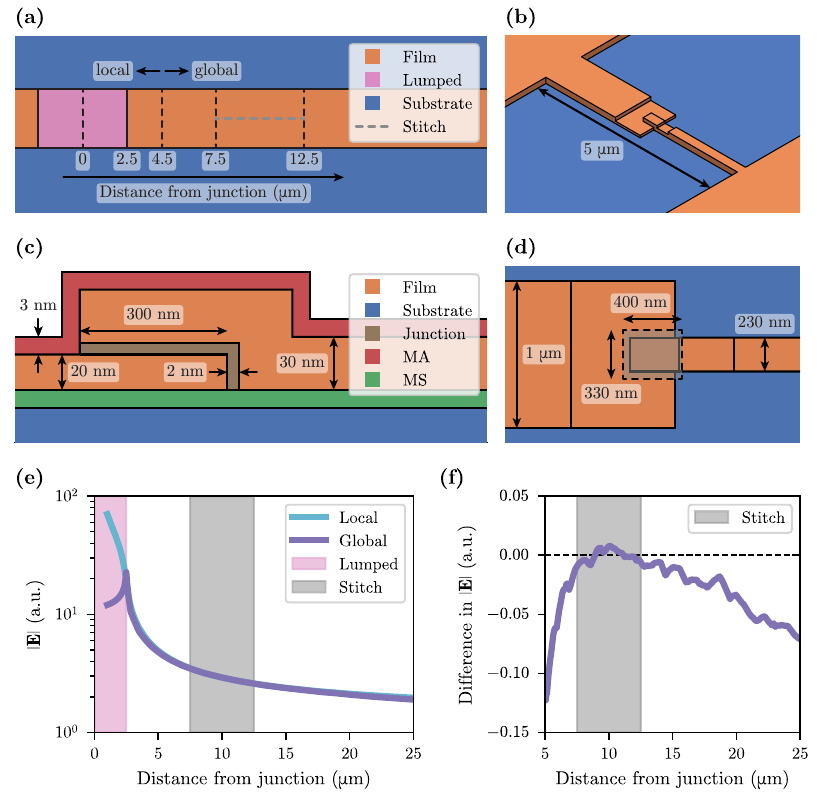}
    \caption{\label{fig:jj} \textbf{Computation of surface participations for the near-junction region.} \textbf{(a)}~Top-down illustration of length scales in the global simulation of the transmon (Ansys HFSS). The near-junction region is modeled as lumped-element within \SI{2.5}{\micro\meter} of the center of the junction. Surface participations are evaluated using the local simulation below \SI{4.5}{\micro\meter} and using the global simulation above \SI{4.5}{\micro\meter}. The stitching line, over which the local-to-global conversion factor is averaged, lies in the plane of the metal-substrate interface between \SI{7.5}{\micro\meter} and \SI{12.5}{\micro\meter}. \textbf{(b)}~Schematic of the near-junction region in the local simulation (Ansys Maxwell 3D), representative of the true geometry realized by the Dolan bridge method. The film thickness is exaggerated for clarity. \textbf{(c)}~Cross-sectional illustration of the junction in the local simulation, not to scale. \textbf{(d)}~Top-down illustration of the junction overlaid with the artificially chosen lossless region, shaded gray. The lossless region has lateral dimensions of \SI{400}{\nano\meter} by \SI{330}{\nano\meter}, compared to \SI{300}{\nano\meter} by \SI{230}{\nano\meter} for the junction. \textbf{(e)}~Magnitude of the electric field $\left|\mathbf{E}\right|$ along the stitching line in the global and local simulations. The scale of $\left|\mathbf{E}\right|$ depends on the simulation settings. In this plot, $\left|\mathbf{E}\right|$ in the local simulation has been multiplied by the local-to-global conversion factor, which is equal to the average ratio of $\left|\mathbf{E}\right|$ in the two simulations along the stitching line (gray). The presence of the lumped element in the global simulation (pink) is responsible for the deviation below \SI{2.5}{\micro\meter}, besides which the two field profiles are close to identical. \textbf{(f)}~Difference in the magnitude of the electric field along the stitching line (global subtract local), after the local-to-global conversion factor has been accounted for. While it is true that evaluating the conversion factor over some stitching region will naturally yield the smallest discrepancy in that region, our chosen stitching region results in the mean discrepancy closest to zero.}
\end{figure*}

A summary of relevant length scales in the global simulation of the transmon is presented in Fig.~\ref{fig:jj}(a). The global simulation includes the film down to within \SI{2.5}{\micro\meter} of the Josephson junction, and evaluates participations down to within \SI{4.5}{\micro\meter} of the junction. The \SI{5}{\micro\meter}-long near-junction region (\SI{\pm2.5}{\micro\meter} either side of the junction) is modeled as a parallel lumped-element circuit with inductance \SI{9}{\nano\henry} and capacitance \SI{6.7}{\femto\farad}, the capacitance having being determined from the local simulation. The pads are drawn as a separate object from the leads, since they receive slightly different donut-to-edge conversion factors. In the 2D simulation, the pads and the leads are drawn as symmetric single films. We exclude the readout resonator and Purcell filter in the global simulation for the purpose of computing surface participations, noting that they contribute very little to the surface participation.

Surface participations in the near-junction region are computed using a 3D electrostatic simulation that resembles the true geometry of our junction, fabricated using the Dolan bridge method, as illustrated in Fig.~\ref{fig:jj}(b). A cross-sectional illustration of the precise geometry of the junction in the local simulation is provided in Fig.~\ref{fig:jj}(c). The local simulation includes the film up to within \SI{50}{\micro\meter} of the junction, although any length scale much larger than the near-junction region would work effectively to negate the electromagnetic influence of the boundary. To complement the global simulation, where participations were evaluated down to within \SI{4.5}{\micro\meter} of the junction, the local simulation evaluates participations up to within \SI{4.5}{\micro\meter} of the junction.

As stated in Ref.~\cite{wang2015}, the local simulation shows that the dominant contribution to surface participations comes from the region immediately next to the junction and from the junction itself. We invoke the same argument that the density of TLSs is low enough that there exists a lossless region around the junction, within which participations are not computed. This argument is supported in experiment by measurements of the merged-element transmon \cite{mamin2021}, whose $Q_\mathrm{int}$ places approximate upper bounds on the loss factor of the near-junction region. In our physical realization of the junction, the electric field rapidly subsides on a length scale of \SI{50}{\nano\meter} outside of the junction, which, in our transmons, have dimensions of \SI{300}{\nano\meter} by \SI{230}{\nano\meter}, and so we choose a lossless region of \SI{400}{\nano\meter} by \SI{330}{\nano\meter}. As shown in Fig.~\ref{fig:jj}(d), the lossless region is centered laterally on the junction.

Unsurprisingly, the total energy stored in the local simulation depends on the excitation voltages imposed by the simulation. To match the energy scales of the local and global simulations, we draw a stitching line that bisects the plane of the MS interface between \SI{7.5}{\micro\meter} and \SI{12.5}{\micro\meter} away from the center of the junction. Figures~\ref{fig:jj}(e),~\ref{fig:jj}(f) confirm that the magnitude of the electric field along the stitching line is close to identical except for a multiplicative local-to-global conversion factor. The participations from the local simulation are therefore multiplied by this local-to-global conversion factor and added to the global participations.

\subsection{Mesh convergence}

\begin{figure*}
    \includegraphics{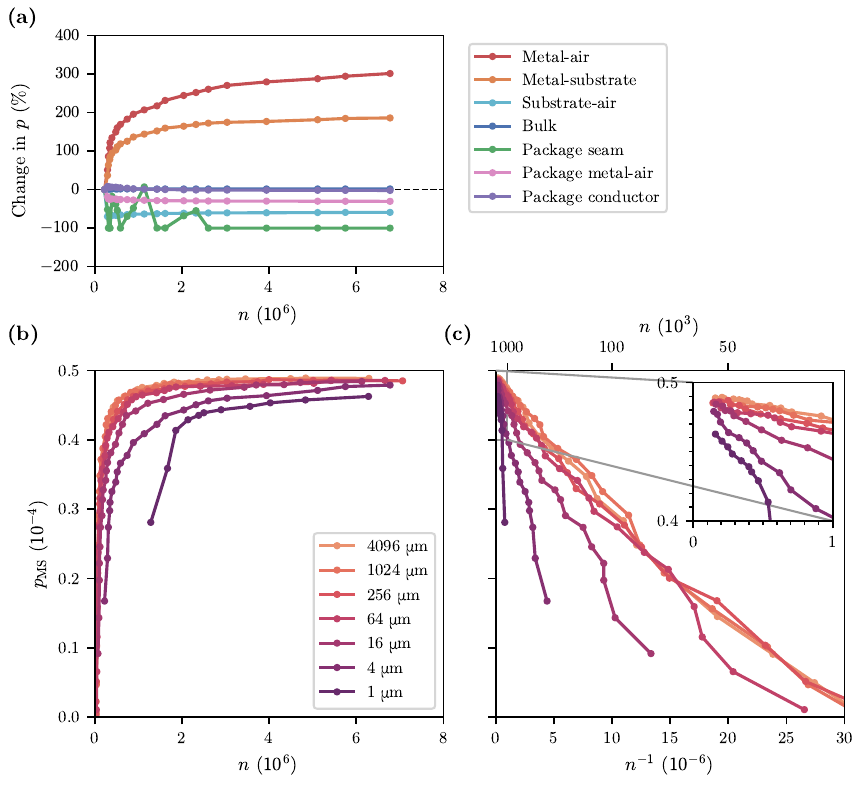}
    \caption{\label{fig:convergence} \textbf{Mesh convergence analysis.} \textbf{(a)}~Participations $p_i$ for the global simulation of the transmon, as a percentage change from those in the first pass, versus number of elements $n$ in a given pass. In this simulation, a maximum element length of \SI{4}{\micro\meter} on the donut region is set as an initial mesh operation. In general, surface participations converge from below and package participations converge from above. In contrast to surface participations, which may change tremendously over the course of the simulation, bulk and package participations converge rapidly. The exception is package seam, which converges poorly due to a deficiency of elements in the region of the seam---a supplementary simulation with different settings is required to obtain package seam reliably. \textbf{(b)}~Metal-substrate (MS) participation $p_\mathrm{MS}$ versus $n$ for varying maximum element length on the donut region, set as an initial mesh operation. \textbf{(c)}~Same as \textbf{(b)}, but $n$ is plotted on a reciprocal scale, such that the $y$-intercept gives the participation in the limit of $n\rightarrow\infty$. The inset, which zooms in to higher $n$, reveals an asymptotic MS participation of \num{4.9e-5}, as recorded in Table~\ref{tab:participations}.}
\end{figure*}

While bulk and package participations tend to converge effortlessly, surface participations may not, and we investigate the reliability of our surface participations by analyzing mesh convergence. Participations are considered to have converged when they do not vary significantly upon further refinement of the mesh. There are two checks that we carry out to evaluate how precisely participations have converged.

Firstly, we plot participations against the number of elements solved in each iteration of the software's adaptive meshing procedure, or ``pass". An example of this is shown in Fig.~\ref{fig:convergence}(a) for the global simulation of the transmon, where a maximum element length of \SI{4}{\micro\meter} on the donut region has been set as an initial mesh operation. We find that surface participations tend to converge from below, while package participations tend to converge from above. We attribute this difference in behavior to the curvature of the geometry over which the participations are being computed---surface participations are computed over a sheet, which has a convex geometry, and so each subsequent pass resolves more and more of the divergence of the electric field around that geometry; conversely, our package has a concave geometry, and so there is no divergence to resolve. The atypical case of package seam will be discussed later in this section.

Secondly, we plot each participation against number of elements for various initial mesh operations, where the operation prescribes a different maximum element length on the donut region. An example of this is shown in Fig.~\ref{fig:convergence}(b) for the case of MS in the global simulation of the transmon. Each trace in this plot represents the convergence of an individual simulation following their respective initial mesh operation. Different traces converge to a similar participation independent of initial mesh as long as the simulation reaches a similar number of elements. In fact, a less stringent initial mesh operation may leave the software to carry out its adaptive meshing procedure more effectively and lead to better convergence.

Even with the resources offered by an HPC cluster, our surface participations are close to converging yet do not converge entirely. Given a simulated system with a number of dimensions $d$, we observe that, as the number of elements $n$ tends to infinity, the characteristic element size $h \propto n^{-1/d}$ tends to zero and the participation $p_n$ approaches some limiting value $p_\infty$, corresponding to the value of $p_n$ in the case of an infinitesimally small mesh size. The asymptotic behavior of this convergence can be approximated by a power law \cite{roache2009}, parameterized by an order of convergence $c$:
\begin{equation}
    p_{\infty}-p_n \propto h^c \propto n^{-\frac{c}{d}}
    \label{eq:convergence}.
\end{equation}

While the convergence of our surface participations is often qualitatively well-described by this relationship, our convergence data are not fit robustly by Eq.~(\ref{eq:convergence}). We speculate that this is related to the nonconstant rate of convergence over the course of the simulation. This is a consequence of the quantity of interest $p$ being a local quantity as opposed to a global quantity, such as frequency, and so in any given adaptive pass, the mesh may or may not target regions that directly influence $p$.

To aid in estimating $p_\infty$, we plot $p_n$ against $n^{-1}$. Extrapolation of $p_n$ to the $y$-intercept, whether by eye or via a linear fit to an appropriate number of the high-$n$ data points, returns the approximate value of $p_\infty$. Only under the special case of $c/d=1$ will this representation yield a linear relationship, but nonetheless it offers a clearer way of viewing convergence behavior. This is shown in Fig.~\ref{fig:convergence}(c) for the case of MS in the global simulation of the transmon, to complement Fig.~\ref{fig:convergence}(b). Convergence of traces with different initial mesh operations towards the same $y$-intercept reinforces our confidence in extrapolating the value of $p_\infty$ in this manner. Critically, we observe that solving for one mode as a setting in the solution setup leads to better convergence for surface participations. On the other hand, solving for more modes, ideally enough to reach package modes, improves reliability of package participations, especially package seam. For this reason, a supplementary simulation that solves for more modes should be carried out to compute package seam participations.

We turn to assessing the impact of participation convergence on extracted loss factors. Although incomplete convergence imparts systematic errors on $\Gamma_i$, such errors need not propagate to the loss budget (and therefore subsequent device design and optimization). Through use of an HPC cluster, our $p_\mathrm{surf}$ for the D1 mode of the tripole stripline resonator reaches a value approximately \SI{21}{\percent} higher than that for the nearly identical design computed in Ref.~\cite{ganjam2024} (``TSLv2"). It follows from the linearity of Eq.~(\ref{eq:matrix}) that, for a fixed $\mathbf{K}$, if a column of $\mathbf{P}$ were multiplied by a given factor, then the corresponding $\Gamma_i$ would be divided by the same factor, and the contribution to the loss budget would remain the same. The requirement that the entire column of $\mathbf{P}$ receives the same multiplicative factor for this case to be valid becomes less consequential the more well-conditioned the matrix is. Crucially, because the D1 mode of the tripole stripline resonator has considerable sensitivity to surface loss, $\Gamma_\mathrm{surf}$ is determined almost entirely by $p_\mathrm{surf}$ of the D1 mode.

To verify this in practice, we construct, in Table~\ref{tab:convergence}, the loss budget for the D1 mode of Tripole~1 using $p_\mathrm{surf}$ computed by this work that uses an HPC, and using $p_\mathrm{surf}$ computed by Ref.~\cite{ganjam2024} that uses a local machine. The relative loss contribution, which for the $i^\mathrm{th}$ loss mechanism is the product of $p_i$ and $\Gamma_i$, is robust against participation scaling. In our case, to correct for a $p_\mathrm{surf}$ that is underestimated by \SI{21}{\percent}, $\Gamma_\mathrm{surf}$ should be reduced by \SI{21}{\percent}.

\begin{table}
    \begin{ruledtabular}
    \begin{tabular}{
        c
        S[table-align-comparator=false,table-format=<2.1]
        S[table-align-comparator=false,table-format=<2.1]
    }
        & \multicolumn{2}{c}{Relative $p_i\Gamma_i$ contribution (\si{\percent})} \\
        \cmidrule{2-3}
        Loss
        & {HPC cluster}
        & {Local machine \cite{ganjam2024}} \\
        \midrule
        Surface & 94.8 & 93.9 \\
        Bulk & 5.1 & 6.0 \\
        Package seam & <0.1 & <0.1 \\
        Package metal-air & <0.1 & <0.1 \\
        Package conductor & <0.1 & <0.1 \\
    \end{tabular}
    \end{ruledtabular}
    \caption{\label{tab:convergence} \textbf{Effect of surface participation convergence.} Each column forms a loss budget for the D1 mode of a representative tripole stripline resonator (``Tripole~1") where surface participations are computed using: (1)~an HPC cluster; (2)~a local machine. The discrepancy in the relative loss contribution by the surface is insignificant.}
\end{table}

\section{LOSS ANALYSIS}\label{appendix:analysis}

The participations for the devices used in this work are listed in Table~\ref{tab:participations}. The results of the loss analyses for individual tripole stripline resonators and segmented stripline resonators are given in Table~\ref{tab:tripole} and Table~\ref{tab:segmented} respectively.

To account for package MA and package conductor in our loss analysis, we use an average of the loss factors extracted by the FWGMR and ellipsoidal cavity in Ref. \cite{lei2023} for 6061 aluminum alloy (``F3", ``F4", ``E2"). The dielectric loss tangent $\tan\delta$ and the surface resistance $R_\mathrm{s}$ are the definitions for the package MA and package conductor loss factors respectively. As two of the data points for $\tan\delta$ are bounds, we take the mean to be mid-range of the upper and lower bounds, and the error as the half-range. The loss factors we adopt are \num{0.084(47)} for package MA and \SI{5.3(1.7)e-6}{\ohm} for package conductor, as listed in Table~\ref{tab:contributions}.

For comparison with tantalum, we use a $\Gamma_\mathrm{surf}$ of \num{4.2(14)e-4} for tantalum on annealed sapphire, extracted by the tripole stripline resonator in Ref.~\cite{ganjam2024}. As explained in Appendix~\ref{appendix:simulation}, we reduce this value by a factor of \SI{21}{\percent}. Similarly, we use a $\Gamma_\mathrm{surf}$ of \num{10.5(2.9)e-4} for aluminum on annealed sapphire and reduce this value by a factor of \SI{21}{\percent}.

\begin{table*}
    \begin{ruledtabular}
    \begin{tabular}{
        c
        S[table-format=1.1e-2]
        S[table-format=1.1e-2]
        S[table-format=1.1e-2]
        S[table-format=1.1e-2]
        S[table-format=1.1e-2]
    }
        & \multicolumn{5}{c}{Participation, $p_i$} \\
        \cmidrule{2-6}
        Loss
        & \text{Tripole (D1)}
        & \text{Tripole (D2)}
        & \text{Tripole (C)}
        & \text{Segmented}
        & \text{Transmon} \\
        \midrule
        Bulk & {0.91} & {0.80} & {0.45} & {0.72} & {0.87} \\
        Re metal-air & 6.7e-5 & 1.4e-6 & 6.3e-7 & 9.1e-6 & 3.8e-6 \\
        Re metal-substrate & 6.4e-4 & 2.4e-5 & 1.3e-5 & 8.5e-5 & 4.9e-5 \\
        Re substrate-air & 7.4e-4 & 2.9e-5 & 1.8e-5 & 9.9e-5 & 5.6e-5 \\
        Al metal-air & {0} & {0} & {0} & 9.1e-6 & 3.8e-6 \\
        Al metal-substrate & {0} & {0} & {0} & 8.5e-5 & 2.0e-5 \\
        Al substrate-air & {0} & {0} & {0} & 9.9e-5 & 2.4e-5 \\
        Re-Al interface (\si{\siemens\per\meter}) & {0} & {0} & {0} & 9.3e4 & 7.1e2 \\
        Package seam (\si{\siemens\per\meter}) & 1.0e-8 & 8.9e-9 & 3.1e-6 & 1.7e-6 & 2.2e-9 \\
        Package metal-air & 8.1e-10 & 2.6e-8 & 1.2e-7 & 4.3e-8 & 4.8e-9 \\
        Package conductor (\si{\per\ohm}) & 3.9e-5 & 1.6e-3 & 4.4e-3 & 1.3e-3 & 4.3e-5 \\
    \end{tabular}
    \end{ruledtabular}
    \caption{\label{tab:participations} \textbf{Participations for the devices used in this work.} Participations are defined and computed using the method described in Appendix~\ref{appendix:simulation}. The tripole and segmented stripline resonators have marginally varying lengths to separate their resonant frequencies, but are assumed to have identical participations. The tripole stripline resonator has no aluminum film, so the corresponding participations are zero. The segmented stripline resonator is assumed to have equal rhenium and aluminum surface participations.}
\end{table*}

\begin{table*}
    \begin{ruledtabular}
    \begin{tabular}{
        c
        S
        S
        S
        S
        S[table-format=1.1(1.1)]
        S[table-format=1.1(1.1)]
        S[table-align-comparator=false, table-format=<2.1(1.1)]
    }
        &
        & \multicolumn{3}{c}{$Q_\mathrm{int}\left(\bar{n}=1\right)$}
        & \multicolumn{3}{c}{Loss factor, $\Gamma_i$} \\
        \cmidrule{3-5}
        \cmidrule{6-8}
        Device
        & {Length (\si{\milli\meter})}
        & {D1 (\num{e6})}
        & {D2 (\num{e6})}
        & \text{C (\num{e6})}
        & {Surface (\num{e-4})}
        & {Bulk (\num{e-8})}
        & {Package seam (\SI{e-3}{\ohm\meter})} \\
        \midrule
        Tripole 1 & 11.95
        & 1.7 & 17.2 & 8.2
        & 3.9(1) & 3.3(5) & 19.6(44) \\
        Tripole 2 & 12.00
        & 1.6 & 21.9 & 15.8
        & 4.3(3) & 1.5(6) & <7.4 \\
        Tripole 3 & 12.05
        & 1.9 & 16.4 & 13.2
        & 3.3(2) & 4.0(5) & <8.7 \\
        Tripole 4 & 12.10
        & 2.2 & 20.1 & 9.2
        & 3.0(1) & 2.9(5) & 16.8(4.6) \\
    \end{tabular}
    \end{ruledtabular}
    \caption{\label{tab:tripole} \textbf{Summary of tripole stripline resonators.} As described in the main text, internal quality factor $Q_\mathrm{int}$ at single-photon power, $\bar{n}=1$, is obtained by interpolation after being fit by a two-level system (TLS) model, and loss factors are extracted by inverting the participation matrix. When the quality of the package seam is sufficiently high, the loss factor can only be bounded. The length refers to the length of the broad strips; the narrow strip receives an extension of \SI{0.5}{\milli\meter}.}
\end{table*}

\begin{table*}
    \begin{ruledtabular}
    \begin{tabular}{
        c
        S[table-format=2.2]
        S[table-format=3]
        S[table-format=1.1]
        S[table-format=1.1(1.1)]
    }
        &
        &
        & {$Q_\mathrm{int}\left(\bar{n}=1\right)$}
        & {Loss factor, $\Gamma_i$} \\
        \cmidrule{4-4}
        \cmidrule{5-5}
        Device
        & {Length (\si{\milli\meter})}
        & {Number of pairs of segments}
        & {(\num{e6})}
        & {Re-Al (\SI{e-12}{\ohm\meter})} \\
        \midrule
        Segmented 1 & 12.46 & 311 & 2.0 & 2.4(9) \\
        Segmented 2 & 12.58 & 314 & 1.7 & 3.5(9) \\
    \end{tabular}
    \end{ruledtabular}
    \caption{\label{tab:segmented} \textbf{Summary of segmented stripline resonators.} As described in the main text, internal quality factor $Q_\mathrm{int}$ at single-photon power, $\bar{n}=1$, is obtained by interpolation after being fit by a two-level system (TLS) model. The loss factor for the rhenium-aluminum interface is extracted by subtracting contributions from all other loss mechanisms and attributing the remainder entirely to that interface.}
\end{table*}

\begin{table*}
    \begin{ruledtabular}
    \begin{tabular}{
        c
        S[table-format=1.2]
        S[table-format=3(2)]
        S[table-format=3(2)]
        S[table-format=3(2)]
    }
         Device
         & {$\omega/2\pi$ (\si{\giga\hertz})}
         & {$T_1$ (\si{\micro\second})}
         & {$T_{2\mathrm{R}}$ (\si{\micro\second})}
         & {$T_{2\mathrm{E}}$ (\si{\micro\second})} \\
         \midrule
         Transmon 1 & 5.00 & 297(22) &  42(19) & 459(33) \\
         Transmon 2 & 5.45 & 277(18) &   21(4) &   51(1) \\
         Transmon 3 & 5.48 & 201(17) & 135(44) & 272(17) \\
         Transmon 4 & 5.26 & 205(40) &  62(24) & 110(12) \\
         Transmon 5 & 5.06 & 407(34) & 410(60) & 656(28) \\
    \end{tabular}
    \end{ruledtabular}
    \caption{\label{tab:transmon} \textbf{Decoherence times of transmons.} Ramsey decoherence time $T_{2\mathrm{R}}$ and the Hahn echo counterpart $T_{2\mathrm{E}}$ for the transmons in Table~\ref{tab:T1} are listed here for completeness. The mean and standard deviation capture temporal fluctuation over the same timescale as for the measurements of relaxation time $T_1$.}
\end{table*}

\clearpage

\bibliography{rhenium}

@article{amin2022,
  title = {Loss Mechanisms in {{TiN}} High Impedance Superconducting Microwave Circuits},
  author = {Amin, Kazi Rafsanjani and Ladner, Carine and Jourdan, Guillaume and Hentz, S{\'e}bastien and Roch, Nicolas and Renard, Julien},
  year = 2022,
  month = apr,
  journal = {Appl. Phys. Lett.},
  volume = {120},
  number = {16},
  pages = {164001},
  doi = {10.1063/5.0086019}
}

@article{arute2019,
  title = {Quantum Supremacy Using a Programmable Superconducting Processor},
  author = {Arute, Frank and Arya, Kunal and Babbush, Ryan and Bacon, Dave and Bardin, Joseph C. and Barends, Rami and Biswas, Rupak and Boixo, Sergio and Brandao, Fernando G. S. L. and Buell, David A. and Burkett, Brian and Chen, Yu and Chen, Zijun and Chiaro, Ben and Collins, Roberto and Courtney, William and Dunsworth, Andrew and Farhi, Edward and Foxen, Brooks and Fowler, Austin and Gidney, Craig and Giustina, Marissa and Graff, Rob and Guerin, Keith and Habegger, Steve and Harrigan, Matthew P. and Hartmann, Michael J. and Ho, Alan and Hoffmann, Markus and Huang, Trent and Humble, Travis S. and Isakov, Sergei V. and Jeffrey, Evan and Jiang, Zhang and Kafri, Dvir and Kechedzhi, Kostyantyn and Kelly, Julian and Klimov, Paul V. and Knysh, Sergey and Korotkov, Alexander and Kostritsa, Fedor and Landhuis, David and Lindmark, Mike and Lucero, Erik and Lyakh, Dmitry and Mandr{\`a}, Salvatore and McClean, Jarrod R. and McEwen, Matthew and Megrant, Anthony and Mi, Xiao and Michielsen, Kristel and Mohseni, Masoud and Mutus, Josh and Naaman, Ofer and Neeley, Matthew and Neill, Charles and Niu, Murphy Yuezhen and Ostby, Eric and Petukhov, Andre and Platt, John C. and Quintana, Chris and Rieffel, Eleanor G. and Roushan, Pedram and Rubin, Nicholas C. and Sank, Daniel and Satzinger, Kevin J. and Smelyanskiy, Vadim and Sung, Kevin J. and Trevithick, Matthew D. and Vainsencher, Amit and Villalonga, Benjamin and White, Theodore and Yao, Z. Jamie and Yeh, Ping and Zalcman, Adam and Neven, Hartmut and Martinis, John M.},
  year = 2019,
  month = oct,
  journal = {Nature},
  volume = {574},
  number = {7779},
  pages = {505--510},
  publisher = {Nature Publishing Group},
  doi = {10.1038/s41586-019-1666-5}
}

@article{baity2024,
  title = {Circle Fit Optimization for Resonator Quality Factor Measurements: {{Point}} Redistribution for Maximal Accuracy},
  shorttitle = {Circle Fit Optimization for Resonator Quality Factor Measurements},
  author = {Baity, Paul G. and Maclean, Connor and Seferai, Valentino and Bronstein, Joe and Shu, Yi and Hemakumara, Tania and Weides, Martin},
  year = 2024,
  month = mar,
  journal = {Phys. Rev. Research},
  volume = {6},
  number = {1},
  pages = {013329},
  doi = {10.1103/PhysRevResearch.6.013329}
}

@article{bal2024,
  title = {Systematic Improvements in Transmon Qubit Coherence Enabled by Niobium Surface Encapsulation},
  author = {Bal, Mustafa and Murthy, Akshay A. and Zhu, Shaojiang and Crisa, Francesco and You, Xinyuan and Huang, Ziwen and Roy, Tanay and Lee, Jaeyel and van Zanten, David and Pilipenko, Roman and Nekrashevich, Ivan and Lunin, Andrei and Bafia, Daniel and Krasnikova, Yulia and Kopas, Cameron J. and Lachman, Ella O. and Miller, Duncan and Mutus, Josh Y. and Reagor, Matthew J. and Cansizoglu, Hilal and Marshall, Jayss and Pappas, David P. and Vu, Kim and Yadavalli, Kameshwar and Oh, Jin-Su and Zhou, Lin and Kramer, Matthew J. and Lecocq, Florent and Goronzy, Dominic P. and {Torres-Castanedo}, Carlos G. and Pritchard, P. Graham and Dravid, Vinayak P. and Rondinelli, James M. and Bedzyk, Michael J. and Hersam, Mark C. and Zasadzinski, John and Koch, Jens and Sauls, James A. and Romanenko, Alexander and Grassellino, Anna},
  year = 2024,
  month = apr,
  journal = {npj Quantum Inf.},
  volume = {10},
  number = {1},
  pages = {43},
  publisher = {Nature Publishing Group},
  doi = {10.1038/s41534-024-00840-x}
}

@article{bejanin2021,
  title = {Interacting Defects Generate Stochastic Fluctuations in Superconducting Qubits},
  author = {B{\'e}janin, J. H. and Earnest, C. T. and Sharafeldin, A. S. and Mariantoni, M.},
  year = 2021,
  month = sep,
  journal = {Phys. Rev. B},
  volume = {104},
  number = {9},
  pages = {094106},
  publisher = {American Physical Society},
  doi = {10.1103/PhysRevB.104.094106}
}

@article{biznarova2024,
  title = {Mitigation of Interfacial Dielectric Loss in Aluminum-on-Silicon Superconducting Qubits},
  author = {Bizn{\'a}rov{\'a}, Janka and Osman, Amr and Rehnman, Emil and Chayanun, Lert and Kri{\v z}an, Christian and Malmberg, Per and Rommel, Marcus and Warren, Christopher and Delsing, Per and Yurgens, August and Bylander, Jonas and Fadavi Roudsari, Anita},
  year = 2024,
  month = aug,
  journal = {npj Quantum Inf.},
  volume = {10},
  number = {1},
  pages = {78},
  publisher = {Nature Publishing Group},
  doi = {10.1038/s41534-024-00868-z}
}

@article{bland2025,
  title = {Millisecond Lifetimes and Coherence Times in {{2D}} Transmon Qubits},
  author = {Bland, Matthew P. and Bahrami, Faranak and Martinez, Jeronimo G. C. and Prestegaard, Paal H. and Smitham, Basil M. and Joshi, Atharv and Hedrick, Elizabeth and Kumar, Shashwat and Yang, Ambrose and {Pakpour-Tabrizi}, Alexander C. and Jindal, Apoorv and Chang, Ray D. and Cheng, Guangming and Yao, Nan and Cava, Robert J. and {de Leon}, Nathalie P. and Houck, Andrew A.},
  year = 2025,
  month = nov,
  journal = {Nature},
  volume = {647},
  number = {8089},
  pages = {343--348},
  publisher = {Nature Publishing Group},
  doi = {10.1038/s41586-025-09687-4}
}

@article{brecht2015,
  title = {Demonstration of Superconducting Micromachined Cavities},
  author = {Brecht, T. and Reagor, M. and Chu, Y. and Pfaff, W. and Wang, C. and Frunzio, L. and Devoret, M. H. and Schoelkopf, R. J.},
  year = 2015,
  month = nov,
  journal = {Appl. Phys. Lett.},
  volume = {107},
  number = {19},
  pages = {192603},
  doi = {10.1063/1.4935541}
}

@article{burnett2019,
  title = {Decoherence Benchmarking of Superconducting Qubits},
  author = {Burnett, Jonathan J. and Bengtsson, Andreas and Scigliuzzo, Marco and Niepce, David and Kudra, Marina and Delsing, Per and Bylander, Jonas},
  year = 2019,
  month = jun,
  journal = {npj Quantum Inf.},
  volume = {5},
  number = {1},
  pages = {54},
  publisher = {Nature Publishing Group},
  doi = {10.1038/s41534-019-0168-5}
}

@article{carroll2022,
  title = {Dynamics of Superconducting Qubit Relaxation Times},
  author = {Carroll, M. and Rosenblatt, S. and Jurcevic, P. and Lauer, I. and Kandala, A.},
  year = 2022,
  month = nov,
  journal = {npj Quantum Inf.},
  volume = {8},
  number = {1},
  pages = {132},
  publisher = {Nature Publishing Group},
  doi = {10.1038/s41534-022-00643-y}
}

@article{chakram2021,
  title = {Seamless {{High-}}\${{Q}}\$ {{Microwave Cavities}} for {{Multimode Circuit Quantum Electrodynamics}}},
  author = {Chakram, Srivatsan and Oriani, Andrew E. and Naik, Ravi K. and Dixit, Akash V. and He, Kevin and Agrawal, Ankur and Kwon, Hyeokshin and Schuster, David I.},
  year = 2021,
  month = aug,
  journal = {Phys. Rev. Lett.},
  volume = {127},
  number = {10},
  pages = {107701},
  publisher = {American Physical Society},
  doi = {10.1103/PhysRevLett.127.107701}
}

@article{chang2013,
  title = {Improved Superconducting Qubit Coherence Using Titanium Nitride},
  author = {Chang, Josephine B. and Vissers, Michael R. and C{\'o}rcoles, Antonio D. and Sandberg, Martin and Gao, Jiansong and Abraham, David W. and Chow, Jerry M. and Gambetta, Jay M. and Beth Rothwell, Mary and Keefe, George A. and Steffen, Matthias and Pappas, David P.},
  year = 2013,
  month = jul,
  journal = {Appl. Phys. Lett.},
  volume = {103},
  number = {1},
  pages = {012602},
  doi = {10.1063/1.4813269}
}

@article{checchin2022,
  title = {Measurement of the {{Low-Temperature Loss Tangent}} of {{High-Resistivity Silicon Using}} a {{High-}}\${{Q}}\$ {{Superconducting Resonator}}},
  author = {Checchin, M. and Frolov, D. and Lunin, A. and Grassellino, A. and Romanenko, A.},
  year = 2022,
  month = sep,
  journal = {Phys. Rev. Appl.},
  volume = {18},
  number = {3},
  pages = {034013},
  publisher = {American Physical Society},
  doi = {10.1103/PhysRevApplied.18.034013}
}

@article{creedon2011,
  title = {High {{Q-factor}} Sapphire Whispering Gallery Mode Microwave Resonator at Single Photon Energies and Millikelvin Temperatures},
  author = {Creedon, Daniel L. and Reshitnyk, Yarema and Farr, Warrick and Martinis, John M. and Duty, Timothy L. and Tobar, Michael E.},
  year = 2011,
  month = jun,
  journal = {Appl. Phys. Lett.},
  volume = {98},
  number = {22},
  pages = {222903},
  doi = {10.1063/1.3595942}
}

@article{crowley2023,
  title = {Disentangling {{Losses}} in {{Tantalum Superconducting Circuits}}},
  author = {Crowley, Kevin D. and McLellan, Russell A. and Dutta, Aveek and Shumiya, Nana and Place, Alexander P. M. and Le, Xuan Hoang and Gang, Youqi and Madhavan, Trisha and Bland, Matthew P. and Chang, Ray and Khedkar, Nishaad and Feng, Yiming Cady and Umbarkar, Esha A. and Gui, Xin and Rodgers, Lila V. H. and Jia, Yichen and Feldman, Mayer M. and Lyon, Stephen A. and Liu, Mingzhao and Cava, Robert J. and Houck, Andrew A. and {de Leon}, Nathalie P.},
  year = 2023,
  month = oct,
  journal = {Phys. Rev. X},
  volume = {13},
  number = {4},
  pages = {041005},
  publisher = {American Physical Society},
  doi = {10.1103/PhysRevX.13.041005}
}

@article{deng2023,
  title = {Titanium {{Nitride Film}} on {{Sapphire Substrate}} with {{Low Dielectric Loss}} for {{Superconducting Qubits}}},
  author = {Deng, Hao and Song, Zhijun and Gao, Ran and Xia, Tian and Bao, Feng and Jiang, Xun and Ku, Hsiang-Sheng and Li, Zhisheng and Ma, Xizheng and Qin, Jin and Sun, Hantao and Tang, Chengchun and Wang, Tenghui and Wu, Feng and Yu, Wenlong and Zhang, Gengyan and Zhang, Xiaohang and Zhou, Jingwei and Zhu, Xing and Shi, Yaoyun and Zhao, Hui-Hai and Deng, Chunqing},
  year = 2023,
  month = feb,
  journal = {Phys. Rev. Appl.},
  volume = {19},
  number = {2},
  pages = {024013},
  publisher = {American Physical Society},
  doi = {10.1103/PhysRevApplied.19.024013}
}

@article{dolan1977,
  title = {Offset Masks for Lift-off Photoprocessing},
  author = {Dolan, G. J.},
  year = 1977,
  month = sep,
  journal = {Appl. Phys. Lett.},
  volume = {31},
  number = {5},
  pages = {337--339},
  doi = {10.1063/1.89690}
}

@article{dumur2016,
  title = {Epitaxial {{Rhenium Microwave Resonators}}},
  author = {Dumur, E. and Delsol, B. and Wei{\ss}l, T. and Kung, B. and Guichard, W. and Hoarau, C. and Naud, C. and Hasselbach, K. and Buisson, O. and Ratter, K. and Gilles, B.},
  year = 2016,
  month = apr,
  journal = {IEEE Trans. Appl. Supercond.},
  volume = {26},
  number = {3},
  pages = {1--4},
  doi = {10.1109/TASC.2016.2547221}
}

@article{ganjam2024,
  title = {Surpassing Millisecond Coherence in on Chip Superconducting Quantum Memories by Optimizing Materials and Circuit Design},
  author = {Ganjam, Suhas and Wang, Yanhao and Lu, Yao and Banerjee, Archan and Lei, Chan U. and Krayzman, Lev and Kisslinger, Kim and Zhou, Chenyu and Li, Ruoshui and Jia, Yichen and Liu, Mingzhao and Frunzio, Luigi and Schoelkopf, Robert J.},
  year = 2024,
  month = may,
  journal = {Nat. Commun.},
  volume = {15},
  number = {1},
  pages = {3687},
  publisher = {Nature Publishing Group},
  doi = {10.1038/s41467-024-47857-6}
}

@phdthesis{ganjam2024a,
  title = {Improving the {{Coherence}} of {{Superconducting Quantum Circuits}} through {{Loss Characterization}} and {{Design Optimization}}},
  author = {Ganjam, Suhas},
  year = 2024,
  month = may,
  school = {Yale University}
}

@phdthesis{gao2008,
  title = {The {{Physics}} of {{Superconducting Microwave Resonators}}},
  author = {Gao, Jiansong},
  year = 2008,
  doi = {10.7907/RAT0-VM75},
  school = {California Institute of Technology}
}

@article{gao2025,
  title = {Establishing a {{New Benchmark}} in {{Quantum Computational Advantage}} with 105-Qubit {{Zuchongzhi}} 3.0 {{Processor}}},
  author = {Gao, Dongxin and Fan, Daojin and Zha, Chen and Bei, Jiahao and Cai, Guoqing and Cai, Jianbin and Cao, Sirui and Chen, Fusheng and Chen, Jiang and Chen, Kefu and Chen, Xiawei and Chen, Xiqing and Chen, Zhe and Chen, Zhiyuan and Chen, Zihua and Chu, Wenhao and Deng, Hui and Deng, Zhibin and Ding, Pei and Ding, Xun and Ding, Zhuzhengqi and Dong, Shuai and Dong, Yupeng and Fan, Bo and Fu, Yuanhao and Gao, Song and Ge, Lei and Gong, Ming and Gui, Jiacheng and Guo, Cheng and Guo, Shaojun and Guo, Xiaoyang and Han, Lianchen and He, Tan and Hong, Linyin and Hu, Yisen and Huang, He-Liang and Huo, Yong-Heng and Jiang, Tao and Jiang, Zuokai and Jin, Honghong and Leng, Yunxiang and Li, Dayu and Li, Dongdong and Li, Fangyu and Li, Jiaqi and Li, Jinjin and Li, Junyan and Li, Junyun and Li, Na and Li, Shaowei and Li, Wei and Li, Yuhuai and Li, Yuan and Liang, Futian and Liang, Xuelian and Liao, Nanxing and Lin, Jin and Lin, Weiping and Liu, Dailin and Liu, Hongxiu and Liu, Maliang and Liu, Xinyu and Liu, Xuemeng and Liu, Yancheng and Lou, Haoxin and Ma, Yuwei and Meng, Lingxin and Mou, Hao and Nan, Kailiang and Nie, Binghan and Nie, Meijuan and Ning, Jie and Niu, Le and Peng, Wenyi and Qian, Haoran and Rong, Hao and Rong, Tao and Shen, Huiyan and Shen, Qiong and Su, Hong and Su, Feifan and Sun, Chenyin and Sun, Liangchao and Sun, Tianzuo and Sun, Yingxiu and Tan, Yimeng and Tan, Jun and Tang, Longyue and Tu, Wenbing and Wan, Cai and Wang, Jiafei and Wang, Biao and Wang, Chang and Wang, Chen and Wang, Chu and Wang, Jian and Wang, Liangyuan and Wang, Rui and Wang, Shengtao and Wang, Xiaomin and Wang, Xinzhe and Wang, Xunxun and Wang, Yeru and Wei, Zuolin and Wei, Jiazhou and Wu, Dachao and Wu, Gang and Wu, Jin and Wu, Shengjie and Wu, Yulin and Xie, Shiyong and Xin, Lianjie and Xu, Yu and Xue, Chun and Yan, Kai and Yang, Weifeng and Yang, Xinpeng and Yang, Yang and Ye, Yangsen and Ye, Zhenping and Ying, Chong and Yu, Jiale and Yu, Qinjing and Yu, Wenhu and Zeng, Xiangdong and Zhan, Shaoyu and Zhang, Feifei and Zhang, Haibin and Zhang, Kaili and Zhang, Pan and Zhang, Wen and Zhang, Yiming and Zhang, Yongzhuo and Zhang, Lixiang and Zhao, Guming and Zhao, Peng and Zhao, Xianhe and Zhao, Xintao and Zhao, Youwei and Zhao, Zhong and Zheng, Luyuan and Zhou, Fei and Zhou, Liang and Zhou, Na and Zhou, Naibin and Zhou, Shifeng and Zhou, Shuang and Zhou, Zhengxiao and Zhu, Chengjun and Zhu, Qingling and Zou, Guihong and Zou, Haonan and Zhang, Qiang and Lu, Chao-Yang and Peng, Cheng-Zhi and Zhu, Xiaobo and Pan, Jian-Wei},
  year = 2025,
  month = mar,
  journal = {Phys. Rev. Lett.},
  volume = {134},
  number = {9},
  pages = {090601},
  publisher = {American Physical Society},
  doi = {10.1103/PhysRevLett.134.090601}
}

@article{googlequantumai2023,
  title = {Suppressing Quantum Errors by Scaling a Surface Code Logical Qubit},
  author = {{Google Quantum AI}},
  year = 2023,
  month = feb,
  journal = {Nature},
  volume = {614},
  number = {7949},
  pages = {676--681},
  publisher = {Nature Publishing Group},
  doi = {10.1038/s41586-022-05434-1}
}

@article{googlequantumaiandcollaborators2025,
  title = {Quantum Error Correction below the Surface Code Threshold},
  author = {{Google Quantum AI and Collaborators}},
  year = 2025,
  month = feb,
  journal = {Nature},
  volume = {638},
  number = {8052},
  pages = {920--926},
  publisher = {Nature Publishing Group},
  doi = {10.1038/s41586-024-08449-y}
}

@article{grunhaupt2018,
  title = {Loss {{Mechanisms}} and {{Quasiparticle Dynamics}} in {{Superconducting Microwave Resonators Made}} of {{Thin-Film Granular Aluminum}}},
  author = {Gr{\"u}nhaupt, Lukas and Maleeva, Nataliya and Skacel, Sebastian T. and Calvo, Martino and {Levy-Bertrand}, Florence and Ustinov, Alexey V. and Rotzinger, Hannes and Monfardini, Alessandro and Catelani, Gianluigi and Pop, Ioan M.},
  year = 2018,
  month = sep,
  journal = {Phys. Rev. Lett.},
  volume = {121},
  number = {11},
  pages = {117001},
  publisher = {American Physical Society},
  doi = {10.1103/PhysRevLett.121.117001}
}

@article{gupta2025,
  title = {Low-Loss Lumped-Element Inductors Made from Granular Aluminum},
  author = {Gupta, Vishakha and Winkel, Patrick and Thakur, Neel and {van Vlaanderen}, Peter and Wang, Yanhao and Ganjam, Suhas and Frunzio, Luigi and Schoelkopf, Robert J.},
  year = 2025,
  month = may,
  journal = {Phys. Rev. Appl.},
  volume = {23},
  number = {5},
  pages = {054067},
  publisher = {American Physical Society},
  doi = {10.1103/PhysRevApplied.23.054067}
}

@article{haq1982,
  title = {Electrical and Superconducting Properties of Rhenium Thin Films},
  author = {Haq, A. Ul and Meyer, O.},
  year = 1982,
  month = aug,
  journal = {Thin Solid Films},
  volume = {94},
  number = {2},
  pages = {119--132},
  doi = {10.1016/0040-6090(82)90504-1}
}

@article{hulm1957,
  title = {Superconducting {{Properties}} of {{Rhenium}}, {{Ruthenium}}, and {{Osmium}}},
  author = {Hulm, J. K. and Goodman, B. B.},
  year = 1957,
  month = may,
  journal = {Phys. Rev.},
  volume = {106},
  number = {4},
  pages = {659--671},
  publisher = {American Physical Society},
  doi = {10.1103/PhysRev.106.659}
}

@article{khalil2012,
  title = {An Analysis Method for Asymmetric Resonator Transmission Applied to Superconducting Devices},
  author = {Khalil, M. S. and Stoutimore, M. J. A. and Wellstood, F. C. and Osborn, K. D.},
  year = 2012,
  month = mar,
  journal = {J. Appl. Phys.},
  volume = {111},
  number = {5},
  pages = {054510},
  doi = {10.1063/1.3692073}
}

@article{klimov2018,
  title = {Fluctuations of {{Energy-Relaxation Times}} in {{Superconducting Qubits}}},
  author = {Klimov, P. V. and Kelly, J. and Chen, Z. and Neeley, M. and Megrant, A. and Burkett, B. and Barends, R. and Arya, K. and Chiaro, B. and Chen, Yu and Dunsworth, A. and Fowler, A. and Foxen, B. and Gidney, C. and Giustina, M. and Graff, R. and Huang, T. and Jeffrey, E. and Lucero, Erik and Mutus, J. Y. and Naaman, O. and Neill, C. and Quintana, C. and Roushan, P. and Sank, Daniel and Vainsencher, A. and Wenner, J. and White, T. C. and Boixo, S. and Babbush, R. and Smelyanskiy, V. N. and Neven, H. and Martinis, John M.},
  year = 2018,
  month = aug,
  journal = {Phys. Rev. Lett.},
  volume = {121},
  number = {9},
  pages = {090502},
  publisher = {American Physical Society},
  doi = {10.1103/PhysRevLett.121.090502}
}

@article{koch2007,
  title = {Charge-Insensitive Qubit Design Derived from the {{Cooper}} Pair Box},
  author = {Koch, Jens and Yu, Terri M. and Gambetta, Jay and Houck, A. A. and Schuster, D. I. and Majer, J. and Blais, Alexandre and Devoret, M. H. and Girvin, S. M. and Schoelkopf, R. J.},
  year = 2007,
  month = oct,
  journal = {Phys. Rev. A},
  volume = {76},
  number = {4},
  pages = {042319},
  publisher = {American Physical Society},
  doi = {10.1103/PhysRevA.76.042319}
}

@article{kono2024,
  title = {Mechanically Induced Correlated Errors on Superconducting Qubits with Relaxation Times Exceeding 0.4 Ms},
  author = {Kono, Shingo and Pan, Jiahe and Chegnizadeh, Mahdi and Wang, Xuxin and Youssefi, Amir and Scigliuzzo, Marco and Kippenberg, Tobias J.},
  year = 2024,
  month = may,
  journal = {Nat. Commun.},
  volume = {15},
  number = {1},
  pages = {3950},
  publisher = {Nature Publishing Group},
  doi = {10.1038/s41467-024-48230-3}
}

@article{krayzman2024,
  title = {Superconducting Quantum Memory with a Suspended Coaxial Resonator},
  author = {Krayzman, Lev and Lei, Chan U and Ganjam, Suhas and Teoh, James and Frunzio, Luigi and Schoelkopf, Robert J.},
  year = 2024,
  month = may,
  journal = {Appl. Phys. Lett.},
  volume = {124},
  number = {20},
  pages = {204001},
  doi = {10.1063/5.0203906}
}

@article{krinner2022,
  title = {Realizing Repeated Quantum Error Correction in a Distance-Three Surface Code},
  author = {Krinner, Sebastian and Lacroix, Nathan and Remm, Ants and Di Paolo, Agustin and Genois, Elie and Leroux, Catherine and Hellings, Christoph and Lazar, Stefania and Swiadek, Francois and Herrmann, Johannes and Norris, Graham J. and Andersen, Christian Kraglund and M{\"u}ller, Markus and Blais, Alexandre and Eichler, Christopher and Wallraff, Andreas},
  year = 2022,
  month = may,
  journal = {Nature},
  volume = {605},
  number = {7911},
  pages = {669--674},
  publisher = {Nature Publishing Group},
  doi = {10.1038/s41586-022-04566-8}
}

@article{lei2020,
  title = {High Coherence Superconducting Microwave Cavities with Indium Bump Bonding},
  author = {Lei, Chan U and Krayzman, Lev and Ganjam, Suhas and Frunzio, Luigi and Schoelkopf, Robert J.},
  year = 2020,
  month = apr,
  journal = {Appl. Phys. Lett.},
  volume = {116},
  number = {15},
  pages = {154002},
  doi = {10.1063/5.0003907}
}

@article{lei2023,
  title = {Characterization of {{Microwave Loss Using Multimode Superconducting Resonators}}},
  author = {Lei, Chan U and Ganjam, Suhas and Krayzman, Lev and Banerjee, Archan and Kisslinger, Kim and Hwang, Sooyeon and Frunzio, Luigi and Schoelkopf, Robert J.},
  year = 2023,
  month = aug,
  journal = {Phys. Rev. Appl.},
  volume = {20},
  number = {2},
  pages = {024045},
  publisher = {American Physical Society},
  doi = {10.1103/PhysRevApplied.20.024045}
}

@article{mamin2021,
  title = {Merged-{{Element Transmons}}: {{Design}} and {{Qubit Performance}}},
  shorttitle = {Merged-{{Element Transmons}}},
  author = {Mamin, H.J. and Huang, E. and Carnevale, S. and Rettner, C.T. and Arellano, N. and Sherwood, M.H. and Kurter, C. and Trimm, B. and Sandberg, M. and Shelby, R.M. and Mueed, M.A. and Madon, B.A. and Pushp, A. and Steffen, M. and Rugar, D.},
  year = 2021,
  month = aug,
  journal = {Phys. Rev. Appl.},
  volume = {16},
  number = {2},
  pages = {024023},
  publisher = {American Physical Society},
  doi = {10.1103/PhysRevApplied.16.024023}
}

@article{martinis2005,
  title = {Decoherence in {{Josephson Qubits}} from {{Dielectric Loss}}},
  author = {Martinis, John M. and Cooper, K. B. and McDermott, R. and Steffen, Matthias and Ansmann, Markus and Osborn, K. D. and Cicak, K. and Oh, Seongshik and Pappas, D. P. and Simmonds, R. W. and Yu, Clare C.},
  year = 2005,
  month = nov,
  journal = {Phys. Rev. Lett.},
  volume = {95},
  number = {21},
  pages = {210503},
  publisher = {American Physical Society},
  doi = {10.1103/PhysRevLett.95.210503}
}

@article{mattis1958,
  title = {Theory of the {{Anomalous Skin Effect}} in {{Normal}} and {{Superconducting Metals}}},
  author = {Mattis, D. C. and Bardeen, J.},
  year = 1958,
  month = jul,
  journal = {Phys. Rev.},
  volume = {111},
  number = {2},
  pages = {412--417},
  publisher = {American Physical Society},
  doi = {10.1103/PhysRev.111.412}
}

@article{mclellan2023,
  title = {Chemical {{Profiles}} of the {{Oxides}} on {{Tantalum}} in {{State}} of the {{Art Superconducting Circuits}}},
  author = {McLellan, Russell A. and Dutta, Aveek and Zhou, Chenyu and Jia, Yichen and Weiland, Conan and Gui, Xin and Place, Alexander P. M. and Crowley, Kevin D. and Le, Xuan Hoang and Madhavan, Trisha and Gang, Youqi and Baker, Lukas and Head, Ashley R. and Waluyo, Iradwikanari and Li, Ruoshui and Kisslinger, Kim and Hunt, Adrian and Jarrige, Ignace and Lyon, Stephen A. and Barbour, Andi M. and Cava, Robert J. and Houck, Andrew A. and Hulbert, Steven L. and Liu, Mingzhao and Walter, Andrew L. and {de Leon}, Nathalie P.},
  year = 2023,
  journal = {Adv. Sci.},
  volume = {10},
  number = {21},
  pages = {2300921},
  doi = {10.1002/advs.202300921}
}

@article{melville2020,
  title = {Comparison of Dielectric Loss in Titanium Nitride and Aluminum Superconducting Resonators},
  author = {Melville, A. and Calusine, G. and Woods, W. and Serniak, K. and Golden, E. and Niedzielski, B. M. and Kim, D. K. and Sevi, A. and Yoder, J. L. and Dauler, E. A. and Oliver, W. D.},
  year = 2020,
  month = sep,
  journal = {Appl. Phys. Lett.},
  volume = {117},
  number = {12},
  pages = {124004},
  doi = {10.1063/5.0021950}
}

@article{milul2023,
  title = {Superconducting {{Cavity Qubit}} with {{Tens}} of {{Milliseconds Single-Photon Coherence Time}}},
  author = {Milul, Ofir and Guttel, Barkay and Goldblatt, Uri and Hazanov, Sergey and Joshi, Lalit M. and Chausovsky, Daniel and Kahn, Nitzan and {\c C}ifty{\"u}rek, Engin and Lafont, Fabien and Rosenblum, Serge},
  year = 2023,
  month = sep,
  journal = {PRX Quantum},
  volume = {4},
  number = {3},
  pages = {030336},
  publisher = {American Physical Society},
  doi = {10.1103/PRXQuantum.4.030336}
}

@article{muller2015,
  title = {Interacting Two-Level Defects as Sources of Fluctuating High-Frequency Noise in Superconducting Circuits},
  author = {M{\"u}ller, Clemens and Lisenfeld, J{\"u}rgen and Shnirman, Alexander and Poletto, Stefano},
  year = 2015,
  month = jul,
  journal = {Phys. Rev. B},
  volume = {92},
  number = {3},
  pages = {035442},
  publisher = {American Physical Society},
  doi = {10.1103/PhysRevB.92.035442}
}

@article{nguyen2019,
  title = {High-{{Coherence Fluxonium Qubit}}},
  author = {Nguyen, Long B. and Lin, Yen-Hsiang and Somoroff, Aaron and Mencia, Raymond and Grabon, Nicholas and Manucharyan, Vladimir E.},
  year = 2019,
  month = nov,
  journal = {Phys. Rev. X},
  volume = {9},
  number = {4},
  pages = {041041},
  publisher = {American Physical Society},
  doi = {10.1103/PhysRevX.9.041041}
}

@book{nielsen2010,
  title = {Quantum {{Computation}} and {{Quantum Information}}: 10th {{Anniversary Edition}}},
  shorttitle = {Quantum {{Computation}} and {{Quantum Information}}},
  author = {Nielsen, Michael A. and Chuang, Isaac L.},
  year = 2010,
  publisher = {Cambridge University Press},
  address = {Cambridge},
  isbn = {978-1-107-00217-3}
}

@article{oconnell2008,
  title = {Microwave Dielectric Loss at Single Photon Energies and Millikelvin Temperatures},
  author = {O'Connell, Aaron D. and Ansmann, M. and Bialczak, R. C. and Hofheinz, M. and Katz, N. and Lucero, Erik and McKenney, C. and Neeley, M. and Wang, H. and Weig, E. M. and Cleland, A. N. and Martinis, J. M.},
  year = 2008,
  month = mar,
  journal = {Appl. Phys. Lett.},
  volume = {92},
  number = {11},
  pages = {112903},
  doi = {10.1063/1.2898887}
}

@article{paik2011,
  title = {Observation of {{High Coherence}} in {{Josephson Junction Qubits Measured}} in a {{Three-Dimensional Circuit QED Architecture}}},
  author = {Paik, Hanhee and Schuster, D. I. and Bishop, Lev S. and Kirchmair, G. and Catelani, G. and Sears, A. P. and Johnson, B. R. and Reagor, M. J. and Frunzio, L. and Glazman, L. I. and Girvin, S. M. and Devoret, M. H. and Schoelkopf, R. J.},
  year = 2011,
  month = dec,
  journal = {Phys. Rev. Lett.},
  volume = {107},
  number = {24},
  pages = {240501},
  publisher = {American Physical Society},
  doi = {10.1103/PhysRevLett.107.240501}
}

@article{pappas2011,
  title = {Two {{Level System Loss}} in {{Superconducting Microwave Resonators}}},
  author = {Pappas, David P. and Vissers, Michael R. and Wisbey, David S. and Kline, Jeffrey S. and Gao, Jiansong},
  year = 2011,
  month = jun,
  journal = {IEEE Trans. Appl. Supercond.},
  volume = {21},
  number = {3},
  pages = {871--874},
  doi = {10.1109/TASC.2010.2097578}
}

@article{place2021,
  title = {New Material Platform for Superconducting Transmon Qubits with Coherence Times Exceeding 0.3 Milliseconds},
  author = {Place, Alexander P. M. and Rodgers, Lila V. H. and Mundada, Pranav and Smitham, Basil M. and Fitzpatrick, Mattias and Leng, Zhaoqi and Premkumar, Anjali and Bryon, Jacob and Vrajitoarea, Andrei and Sussman, Sara and Cheng, Guangming and Madhavan, Trisha and Babla, Harshvardhan K. and Le, Xuan Hoang and Gang, Youqi and J{\"a}ck, Berthold and Gyenis, Andr{\'a}s and Yao, Nan and Cava, Robert J. and {de Leon}, Nathalie P. and Houck, Andrew A.},
  year = 2021,
  month = mar,
  journal = {Nat. Commun.},
  volume = {12},
  number = {1},
  pages = {1779},
  publisher = {Nature Publishing Group},
  doi = {10.1038/s41467-021-22030-5}
}

@article{potts2001,
  title = {Novel Fabrication Methods for Submicrometer {{Josephson}} Junction Qubits},
  author = {Potts, A. and Routley, P. R. and Parker, G. J. and Baumberg, J. J. and {de Groot}, P. A. J.},
  year = 2001,
  month = jun,
  journal = {J. Mater. Sci.: Mater. Electron.},
  volume = {12},
  number = {4},
  pages = {289--293},
  doi = {10.1023/A:1011279908265}
}

@article{probst2015,
  title = {Efficient and Robust Analysis of Complex Scattering Data under Noise in Microwave Resonators},
  author = {Probst, S. and Song, F. B. and Bushev, P. A. and Ustinov, A. V. and Weides, M.},
  year = 2015,
  month = feb,
  journal = {Rev. Sci. Instrum.},
  volume = {86},
  number = {2},
  pages = {024706},
  doi = {10.1063/1.4907935}
}

@phdthesis{ratter2017,
  title = {Epitaxial {{Rhenium}}, a Clean Limit Superconductor for Superconducting {{Qbits}}},
  author = {Ratter, Kitti},
  year = 2017,
  month = oct,
  number = {2017GREAY074},
  school = {Universit\'e Grenoble Alpes}
}

@article{read2023,
  title = {Precision {{Measurement}} of the {{Microwave Dielectric Loss}} of {{Sapphire}} in the {{Quantum Regime}} with {{Parts-per-Billion Sensitivity}}},
  author = {Read, Alexander P. and Chapman, Benjamin J. and Lei, Chan U and Curtis, Jacob C. and Ganjam, Suhas and Krayzman, Lev and Frunzio, Luigi and Schoelkopf, Robert J.},
  year = 2023,
  month = mar,
  journal = {Phys. Rev. Appl.},
  volume = {19},
  number = {3},
  pages = {034064},
  publisher = {American Physical Society},
  doi = {10.1103/PhysRevApplied.19.034064}
}

@article{reagor2016,
  title = {Quantum Memory with Millisecond Coherence in Circuit {{QED}}},
  author = {Reagor, Matthew and Pfaff, Wolfgang and Axline, Christopher and Heeres, Reinier W. and Ofek, Nissim and Sliwa, Katrina and Holland, Eric and Wang, Chen and Blumoff, Jacob and Chou, Kevin and Hatridge, Michael J. and Frunzio, Luigi and Devoret, Michel H. and Jiang, Liang and Schoelkopf, Robert J.},
  year = 2016,
  month = jul,
  journal = {Phys. Rev. B},
  volume = {94},
  number = {1},
  pages = {014506},
  publisher = {American Physical Society},
  doi = {10.1103/PhysRevB.94.014506}
}

@article{rieger2023,
  title = {Fano {{Interference}} in {{Microwave Resonator Measurements}}},
  author = {Rieger, D. and G{\"u}nzler, S. and Spiecker, M. and Nambisan, A. and Wernsdorfer, W. and Pop, I.M.},
  year = 2023,
  month = jul,
  journal = {Phys. Rev. Appl.},
  volume = {20},
  number = {1},
  pages = {014059},
  publisher = {American Physical Society},
  doi = {10.1103/PhysRevApplied.20.014059}
}

@book{roache2009,
  title = {Fundamentals of Verification and Validation},
  author = {Roache, Patrick J.},
  year = 2009,
  publisher = {Hermosa Publishers},
  address = {Socorro, NM},
  isbn = {978-0-913478-12-7}
}

@article{sage2011,
  title = {Study of Loss in Superconducting Coplanar Waveguide Resonators},
  author = {Sage, Jeremy M. and Bolkhovsky, Vladimir and Oliver, William D. and Turek, Benjamin and Welander, Paul B.},
  year = 2011,
  month = mar,
  journal = {J. Appl. Phys.},
  volume = {109},
  number = {6},
  pages = {063915},
  doi = {10.1063/1.3552890}
}

@article{schlor2019,
  title = {Correlating {{Decoherence}} in {{Transmon Qubits}}: {{Low Frequency Noise}} by {{Single Fluctuators}}},
  shorttitle = {Correlating {{Decoherence}} in {{Transmon Qubits}}},
  author = {Schl{\"o}r, Steffen and Lisenfeld, J{\"u}rgen and M{\"u}ller, Clemens and Bilmes, Alexander and Schneider, Andre and Pappas, David P. and Ustinov, Alexey V. and Weides, Martin},
  year = 2019,
  month = nov,
  journal = {Phys. Rev. Lett.},
  volume = {123},
  number = {19},
  pages = {190502},
  publisher = {American Physical Society},
  doi = {10.1103/PhysRevLett.123.190502}
}

@article{schreier2008,
  title = {Suppressing Charge Noise Decoherence in Superconducting Charge Qubits},
  author = {Schreier, J. A. and Houck, A. A. and Koch, Jens and Schuster, D. I. and Johnson, B. R. and Chow, J. M. and Gambetta, J. M. and Majer, J. and Frunzio, L. and Devoret, M. H. and Girvin, S. M. and Schoelkopf, R. J.},
  year = 2008,
  month = may,
  journal = {Phys. Rev. B},
  volume = {77},
  number = {18},
  pages = {180502},
  publisher = {American Physical Society},
  doi = {10.1103/PhysRevB.77.180502}
}

@article{sivak2023,
  title = {Real-Time Quantum Error Correction beyond Break-Even},
  author = {Sivak, V. V. and Eickbusch, A. and Royer, B. and Singh, S. and Tsioutsios, I. and Ganjam, S. and Miano, A. and Brock, B. L. and Ding, A. Z. and Frunzio, L. and Girvin, S. M. and Schoelkopf, R. J. and Devoret, M. H.},
  year = 2023,
  month = apr,
  journal = {Nature},
  volume = {616},
  number = {7955},
  pages = {50--55},
  publisher = {Nature Publishing Group},
  doi = {10.1038/s41586-023-05782-6}
}

@article{song2009,
  title = {Microwave Response of Vortices in Superconducting Thin Films of {{Re}} and {{Al}}},
  author = {Song, C. and Heitmann, T. W. and DeFeo, M. P. and Yu, K. and McDermott, R. and Neeley, M. and Martinis, John M. and Plourde, B. L. T.},
  year = 2009,
  month = may,
  journal = {Phys. Rev. B},
  volume = {79},
  number = {17},
  pages = {174512},
  publisher = {American Physical Society},
  doi = {10.1103/PhysRevB.79.174512}
}

@article{tarkaeva2025,
  title = {High-Performance Amorphous Superconducting Rhenium Films by e-Beam Evaporation},
  author = {Tarkaeva, E. V. and Ievleva, V. A. and Prishchepa, A. R. and Zhukova, E. S. and Terentiev, A. and Kuntsevich, A. {\relax Yu}.},
  year = 2025,
  month = sep,
  journal = {J. Appl. Phys.},
  volume = {138},
  number = {12},
  pages = {123904},
  doi = {10.1063/5.0292196}
}

@article{teknowijoyo2023,
  title = {Superconducting {{Polycrystalline Rhenium Films Deposited}} at {{Room Temperature}}},
  author = {Teknowijoyo, S. and Gulian, A.},
  year = 2023,
  month = dec,
  journal = {Opt. Mem. Neural Netw.},
  volume = {32},
  number = {3},
  pages = {S327-S333},
  doi = {10.3103/S1060992X23070184}
}

@article{tuokkola2025,
  title = {Methods to Achieve Near-Millisecond Energy Relaxation and Dephasing Times for a Superconducting Transmon Qubit},
  author = {Tuokkola, Mikko and Sunada, Yoshiki and Kivij{\"a}rvi, Heidi and Albanese, Jonatan and Gr{\"o}nberg, Leif and Kaikkonen, Jukka-Pekka and Vesterinen, Visa and Govenius, Joonas and M{\"o}tt{\"o}nen, Mikko},
  year = 2025,
  month = jul,
  journal = {Nat. Commun.},
  volume = {16},
  number = {1},
  pages = {5421},
  publisher = {Nature Publishing Group},
  doi = {10.1038/s41467-025-61126-0}
}

@article{wang2009,
  title = {Improving the Coherence Time of Superconducting Coplanar Resonators},
  author = {Wang, H. and Hofheinz, M. and Wenner, J. and Ansmann, M. and Bialczak, R. C. and Lenander, M. and Lucero, Erik and Neeley, M. and O'Connell, A. D. and Sank, D. and Weides, M. and Cleland, A. N. and Martinis, John M.},
  year = 2009,
  month = dec,
  journal = {Appl. Phys. Lett.},
  volume = {95},
  number = {23},
  pages = {233508},
  doi = {10.1063/1.3273372}
}

@article{wang2015,
  title = {Surface Participation and Dielectric Loss in Superconducting Qubits},
  author = {Wang, C. and Axline, C. and Gao, Y. Y. and Brecht, T. and Chu, Y. and Frunzio, L. and Devoret, M. H. and Schoelkopf, R. J.},
  year = 2015,
  month = oct,
  journal = {Appl. Phys. Lett.},
  volume = {107},
  number = {16},
  pages = {162601},
  doi = {10.1063/1.4934486}
}

@article{wang2022,
  title = {Towards Practical Quantum Computers: Transmon Qubit with a Lifetime Approaching 0.5 Milliseconds},
  shorttitle = {Towards Practical Quantum Computers},
  author = {Wang, Chenlu and Li, Xuegang and Xu, Huikai and Li, Zhiyuan and Wang, Junhua and Yang, Zhen and Mi, Zhenyu and Liang, Xuehui and Su, Tang and Yang, Chuhong and Wang, Guangyue and Wang, Wenyan and Li, Yongchao and Chen, Mo and Li, Chengyao and Linghu, Kehuan and Han, Jiaxiu and Zhang, Yingshan and Feng, Yulong and Song, Yu and Ma, Teng and Zhang, Jingning and Wang, Ruixia and Zhao, Peng and Liu, Weiyang and Xue, Guangming and Jin, Yirong and Yu, Haifeng},
  year = 2022,
  month = jan,
  journal = {npj Quantum Inf.},
  volume = {8},
  number = {1},
  pages = {3},
  publisher = {Nature Publishing Group},
  doi = {10.1038/s41534-021-00510-2}
}

@article{wenner2011,
  title = {Surface Loss Simulations of Superconducting Coplanar Waveguide Resonators},
  author = {Wenner, J. and Barends, R. and Bialczak, R. C. and Chen, Yu and Kelly, J. and Lucero, Erik and Mariantoni, Matteo and Megrant, A. and O'Malley, P. J. J. and Sank, D. and Vainsencher, A. and Wang, H. and White, T. C. and Yin, Y. and Zhao, J. and Cleland, A. N. and Martinis, John M.},
  year = 2011,
  month = sep,
  journal = {Appl. Phys. Lett.},
  volume = {99},
  number = {11},
  pages = {113513},
  doi = {10.1063/1.3637047}
}

@article{woods2019,
  title = {Determining {{Interface Dielectric Losses}} in {{Superconducting Coplanar-Waveguide Resonators}}},
  author = {Woods, W. and Calusine, G. and Melville, A. and Sevi, A. and Golden, E. and Kim, D.K. and Rosenberg, D. and Yoder, J.L. and Oliver, W.D.},
  year = 2019,
  month = jul,
  journal = {Phys. Rev. Appl.},
  volume = {12},
  number = {1},
  pages = {014012},
  publisher = {American Physical Society},
  doi = {10.1103/PhysRevApplied.12.014012}
}

@article{wu2021,
  title = {Strong {{Quantum Computational Advantage Using}} a {{Superconducting Quantum Processor}}},
  author = {Wu, Yulin and Bao, Wan-Su and Cao, Sirui and Chen, Fusheng and Chen, Ming-Cheng and Chen, Xiawei and Chung, Tung-Hsun and Deng, Hui and Du, Yajie and Fan, Daojin and Gong, Ming and Guo, Cheng and Guo, Chu and Guo, Shaojun and Han, Lianchen and Hong, Linyin and Huang, He-Liang and Huo, Yong-Heng and Li, Liping and Li, Na and Li, Shaowei and Li, Yuan and Liang, Futian and Lin, Chun and Lin, Jin and Qian, Haoran and Qiao, Dan and Rong, Hao and Su, Hong and Sun, Lihua and Wang, Liangyuan and Wang, Shiyu and Wu, Dachao and Xu, Yu and Yan, Kai and Yang, Weifeng and Yang, Yang and Ye, Yangsen and Yin, Jianghan and Ying, Chong and Yu, Jiale and Zha, Chen and Zhang, Cha and Zhang, Haibin and Zhang, Kaili and Zhang, Yiming and Zhao, Han and Zhao, Youwei and Zhou, Liang and Zhu, Qingling and Lu, Chao-Yang and Peng, Cheng-Zhi and Zhu, Xiaobo and Pan, Jian-Wei},
  year = 2021,
  month = oct,
  journal = {Phys. Rev. Lett.},
  volume = {127},
  number = {18},
  pages = {180501},
  publisher = {American Physical Society},
  doi = {10.1103/PhysRevLett.127.180501}
}

@article{zhang2024,
  title = {Acceptor-{{Induced Bulk Dielectric Loss}} in {{Superconducting Circuits}} on {{Silicon}}},
  author = {Zhang, Zi-Huai and Godeneli, Kadircan and He, Justin and Odeh, Mutasem and Zhou, Haoxin and Meesala, Srujan and Sipahigil, Alp},
  year = 2024,
  month = oct,
  journal = {Phys. Rev. X},
  volume = {14},
  number = {4},
  pages = {041022},
  publisher = {American Physical Society},
  doi = {10.1103/PhysRevX.14.041022}
}

@article{zhu2022,
  title = {Quantum Computational Advantage via 60-Qubit 24-Cycle Random Circuit Sampling},
  author = {Zhu, Qingling and Cao, Sirui and Chen, Fusheng and Chen, Ming-Cheng and Chen, Xiawei and Chung, Tung-Hsun and Deng, Hui and Du, Yajie and Fan, Daojin and Gong, Ming and Guo, Cheng and Guo, Chu and Guo, Shaojun and Han, Lianchen and Hong, Linyin and Huang, He-Liang and Huo, Yong-Heng and Li, Liping and Li, Na and Li, Shaowei and Li, Yuan and Liang, Futian and Lin, Chun and Lin, Jin and Qian, Haoran and Qiao, Dan and Rong, Hao and Su, Hong and Sun, Lihua and Wang, Liangyuan and Wang, Shiyu and Wu, Dachao and Wu, Yulin and Xu, Yu and Yan, Kai and Yang, Weifeng and Yang, Yang and Ye, Yangsen and Yin, Jianghan and Ying, Chong and Yu, Jiale and Zha, Chen and Zhang, Cha and Zhang, Haibin and Zhang, Kaili and Zhang, Yiming and Zhao, Han and Zhao, Youwei and Zhou, Liang and Lu, Chao-Yang and Peng, Cheng-Zhi and Zhu, Xiaobo and Pan, Jian-Wei},
  year = 2022,
  month = feb,
  journal = {Sci. Bull.},
  volume = {67},
  number = {3},
  pages = {240--245},
  doi = {10.1016/j.scib.2021.10.017}
}

\end{document}